\documentclass[12pt]{article}

\def\bmtht{\bm\theta}
\usepackage{bm}
\usepackage{natbib}
\usepackage{amsmath,amsthm,epsfig,amsfonts,color}
\usepackage{multirow}
\usepackage{floatrow}
\newfloatcommand{capbtabbox}{table}[][\FBwidth]
\usepackage{rotating}
\usepackage{subfigure}
\usepackage{verbatim}
\usepackage{pdfpages}
\usepackage{threeparttable}
\usepackage{graphics, graphicx, hyperref, url, amsmath, subfigure, amsthm, amsfonts, enumerate, epstopdf, caption, float, booktabs}
\usepackage{comment}
\newtheorem{theorem}{{Theorem}}
\newtheorem{lemma}{{Lemma}}
\newtheorem{remark}{{Remark}}
\renewcommand{\raggedright}{\leftskip=0pt \rightskip=0pt plus 0cm}
\def\beq{\begin{equation}}
\def\eeq{\end{equation}}
\def\beqr{\begin{eqnarray}}
\def\eeqr{\end{eqnarray}}
\def\beqrs{\begin{eqnarray*}}
\def\eeqrs{\end{eqnarray*}}
\def\bet{\begin{theorem}}
\def\eet{\end{theorem}}
\def\bel{\begin{lemma}}
\def\eel{\end{lemma}}
\def\bep{\begin{proposition}}
\def\eep{\end{proposition}}
\def\bec{\begin{corollary}}
\def\eec{\end{corollary}}

\begin{document}

  \title{Empirical likelihood-based portmanteau tests for autoregressive moving average models with possible infinite variance innovation}
\author{Xiaohui Liu, Donghui Fan, Xu Zhang\thanks{
Xiaohui Liu is Professor at School of Statistics, and Key Laboratory of Data Science in Finance and Economics, Jiangxi University of Finance and Economic, Nanchang, China. E-mail: liuxiaohui@jxufe.edu.cn.
Donghui Fan is a graduate student at School of Statistics, Jiangxi University of Finance and Economic, Nanchang, China. E-mail: donghui.fan@jxufe.edu.cn.
Xu Zhang is Postdoctoral Fellow at School of Mathematical Science, South China Normal University, Guangzhou, China. E-mail: zhangx6690@gmail.com.
Catherine C. Liu is Associate Professor at Department of Applied Mathematics, The Hong Kong Polytechnic University, Hong Kong SAR.
E-mail: macliu@polyu.edu.hk.
Xu Zhang is the corresponding author.}
~and Catherine C. Liu}
  \maketitle

\begin{abstract}
	It is an important task in the literature to check whether a fitted autoregressive moving average (ARMA) model is adequate, while the currently used tests may suffer from the size distortion problem when the underlying autoregressive models have low persistence. To fill this gap, this paper proposes two empirical likelihood-based portmanteau tests.
	The first one is naive but can serve as a benchmark, and the second is for the case with infinite variance innovations. The asymptotic distributions under the null hypothesis are derived under mild moment conditions, and their usefulness is demonstrated by simulation experiments and two real data examples.
\end{abstract}

{\it Keywords: ARMA model; GARCH process; diagnostic checking; empirical likelihood; infinite variance}

\section{Introduction}
\label{Introduction}

Consider the autoregressive moving average (ARMA) model with orders $p$ and $q$, denoted by ARMA($p, q$),
\begin{eqnarray}\label{eqn:model}
  X_t = \mu + \sum_{i=1}^p \phi_i X_{t-i} + \sum_{j=1}^q \psi_j \varepsilon_{t-j} + \varepsilon_{t},
\end{eqnarray}
where $(\mu, \phi_1, \cdots, \phi_p, \psi_1, \cdots, \psi_q)$ contains unknown parameters, and $\{\varepsilon_{t}\}$ is a martingale difference series, and this model has been widely used in many fields such as finance and economics. It is an important task in time series analysis to check whether the fitted model is adequate, i.e. the orders $p$ or/and $q$ may not be correctly specified, and there is a huge literature for it.
The early seminal works include the Box-Pierce statistic $Q_m$ in \cite{BoxPierce1970JASA} and Ljung-Box statistic $\tilde Q_m$ in \cite{LjungBox1978Biometrika}, and they can be defined as
\begin{eqnarray*}
  Q_m = n \hat{\bm\rho}^\top \hat{\bm\rho}, \quad \text{and} \quad \tilde Q_m = n \hat{\bm\rho}^\top W \hat{\bm\rho},
\end{eqnarray*}
where the diagonal matrix $W = \text{diag}\{(n+2)/(n-1), (n+2)/(n-2), \cdots, (n+2)/(n-m)\}$, $\hat{\bm\rho} := (\hat\rho_1, \hat\rho_2, \cdots, \hat\rho_m)^\top$, the residuals auto-correlation at lag $k$ has the form of
\begin{eqnarray*}
  \hat\rho_k = \sum_{t=k+1}^n \hat{\varepsilon}_t \hat{\varepsilon}_{t-k} / \sum_{t=1}^n \hat{\varepsilon}_t^2,
\end{eqnarray*}
and $\{\hat{\varepsilon}_t\}$ are residuals from the fitted ARMA model at \eqref{eqn:model}.

Note that $\tilde Q_m$ is a weighted version of $Q_m$, and it usually has a better performance especially when the sample size $n$ is relatively small. Further improvements along this line include the weighted Ljund-Box test statistic in \cite{FisherGallagher2012JASA}. All these test statistics are easily implemented, and hence they have already been widely applied in practice. However, as pointed out by \cite{zhu2016bootstrapping}, the asymptotic properties of these test statistics are only valid under a strong condition that $\{\varepsilon_t\}$ are independent and identically distributed ($i.i.d.$) random variables.
\cite{zhu2016bootstrapping} developed an interesting random weighting (RW) technique to calculate the critical values of these test statistics, and hence these easy-to-implemented tests can be extended to the case that $\varepsilon_t$'s are uncorrelated, but not necessarily independent.

In the meanwhile, when autoregressive (AR) models have low persistence, i.e., AR coefficients are relatively small, the RW method in \cite{zhu2016bootstrapping} still suffers from significant size distortion; see simulation results in Section 3 for details.
As a result, this paper revisits the literature of diagnostic checking for AMRA models, and a new test statistic is then proposed by the profile empirical likelihood (EL) method \citep{owen2001empirical, qin1994empirical}. It can be further shown that, under mild conditions, the proposed test statistic has the null chi-squared distribution, which is a desirable property for tests.

On the other hand, financial and economic data usually exhibit the phenomenon of volatility clustering, which can be interpreted by the conditional heteroscedasticity.
\cite{engle1982autoregressive} first suggested an autoregressive conditional heteroskedastic (ARCH) model for it. Moreover, by noting that the AR process usually needs a higher order than the ARMA process in the actual modeling, \cite{Bollerslev1994review} extended the ARCH model to a more flexible generalized autoregressive conditional heteroskedastic (GARCH) model, which not only reduces the number of parameters but also provides a better fit to the data; see, e.g., \cite{mikosch2000limit}, \cite{hall2003inference}, \cite{peng2003least}, \cite{chan2010inference}, \cite{ling2007self}, \cite{ma2021test}, and references therein.
The GARCH model has the form of
\begin{eqnarray}\label{eqn:errors}
	\varepsilon_{t} = \eta_{t} \sigma_{t}, ~ \sigma_{t}^2 = \omega + \sum_{i=1}^r a_i  \varepsilon_{t-i}^2 +  \sum_{j=1}^s b_j   \sigma_{t-j}^2,
\end{eqnarray}
where $\{\eta_{t}\}$ are $i.i.d.$ random errors with means zero and variances one, $(\omega, a_1, \cdots, a_r, b_1, \cdots,\\$ $ b_s)$ contains unknown parameters, and $\omega$, $a_i$'s and $b_j$'s are assumed to be positive. The GARCH process has the finite variance if $\sum_{i=1}^r a_i+\sum_{j=1}^s b_j<1$, however, many financial data may exhibit an infinite variance of $\{\varepsilon_{t}\}$, i.e. $\sum_{i=1}^r a_i+\sum_{j=1}^s b_j$ may be very close to one.

When $\varepsilon_{t}$ has an infinite variance, both the RW and EL test statistics perform poorly in terms of both sizes and powers since they are only valid for the case with finite variance innovations.
As a result, this paper further proposes a weighted empirical likelihood (WeL) test statistic to check the adequacy of the fitted models at \eqref{eqn:model} with GARCH errors at \eqref{eqn:errors}, and the null distribution is also derived with the innovations being allowed to have infinite variance.

Both EL and WeL are developed based on the empirical likelihood methods in \cite{qin1994empirical}, and the original is attributed to \cite{owen2001empirical}.
Empirical likelihood is a popular nonparametric likelihood method and has wide and successful applications in many fields; see \cite{shen2016empirical}, \cite{shen2019semiparametric} and among others.
While it has attracted less attention in literature of time series. Empirical likelihood is first introduced by \cite{ChanLing2006ET} to GARCH models to build likelihood ratio test statistics, and other applications include but are limited to constructing confidence intervals for the tail index and testing for zero median of errors; see \cite{ZhangLiPeng2019}, \cite{ma2021test}, etc.
It is noteworthy to point out that the naive EL test only works for finite variance innovations, while the proposed WeL method is motivated by the self-weighting method to modify local quasi-maximum likelihood estimators in\cite{ling2007self}.

The remainder of the paper is organized as follows. Section 2 gives two tests, and their null distributions are also derived. Sections 3 and 4 provide simulation results and real analysis, respectively, and a quick summary is given in Section 5. The theoretical details are relegated to the Appendix.

\vskip 0.1 in
\section{Methodology and main results}
\label{Methodology}

Let $\bm\gamma := (\gamma_1, \gamma_2, \cdots, \gamma_m)^\top = (E(\varepsilon_t \varepsilon_{t-1}), E(\varepsilon_t \varepsilon_{t-2}), \cdots, $ $E(\varepsilon_t \varepsilon_{t-m}))$ for some given $m \ge 1$, and $\bm\theta = (\mu, \phi_1, \cdots, \phi_p$, $\psi_1, \cdots, \psi_q)^\top$. The serial correlation hypotheses can be summarized into
\begin{eqnarray}\label{eqn:H0}
  \mathcal{H}_0: \bm \gamma = 0 \quad \text{versus} \quad \mathcal{H}_1: \bm \gamma \neq 0.
\end{eqnarray}

Assume that the observed time series $\{X_t\}_{t=1}^n$ are generated from model \eqref{eqn:model}. Note that the definition of $\bm\gamma$ is related to the expectation. We propose to test \eqref{eqn:H0} by using the empirical likelihood technique in \cite{qin1994empirical}. We start with the case that $\varepsilon_t$ has finite variance, and then extend the result to the weighted empirical likelihood test statistic for infinite variance innovations.

\subsection{Finite variance innovations}
For convenience, define $\varepsilon_{t}(\bm\theta) = X_t - \mu - \sum_{i=1}^p\phi_i X_{t-i} - \sum_{j=1}^q \psi_j \varepsilon_{t-j}(\bm\theta)$. Note that the least squares (LS) estimator $\hat{\bm\theta}$ minimizes
\begin{eqnarray*}
  \sum_{t=1}^n \left(X_t - \mu - \sum_{i=1}^p\phi_i X_{t-i} - \sum_{j=1}^q \psi_j \varepsilon_{t-j}(\bm\theta)\right)^2.
\end{eqnarray*}
That is, $\hat{\bm\theta}$ is the solution to
\begin{eqnarray}\label{eqn:equation}
  \sum_{t=1}^n \varepsilon_{t}(\bm\theta) \frac{\partial \varepsilon_{t}(\bm\theta)}{\partial \bm\theta} = 0.
\end{eqnarray}
This motivates us to define the empirical likelihood function for testing $\mathcal{H}_0$ as follows:
\begin{eqnarray*}
  L(\bmtht, \bm \gamma) &= \sup \{{\prod_{t=m+1}^n}(Np_t): p_{m+1} \geq 0, \cdots, p_n \geq 0, \\\nonumber
  &\sum_{t=m+1}^n p_t=1, \sum_{t=m+1}^n p_{t} \boldsymbol Z_{t}(\bmtht, \bm\gamma)=0\},
\end{eqnarray*}
where $N = n - m$, and $ \boldsymbol Z_{t}(\bm\theta, \bm \gamma) = (Z_{t,1}(\bmtht, \bm\gamma)^\top,$ $ Z_{t,p+q+1}(\bmtht, \bm \gamma), \cdots, Z_{t,p+q+m}(\bmtht, \bm \gamma))^\top$ with
\begin{eqnarray*}
  \begin{cases}
\bm Z_{t, 1}(\bmtht, \bm\gamma) = \varepsilon_{t}(\bm\theta)\frac{\partial \varepsilon_{t}(\bm\theta)}{\partial \bm\theta}, \\[2ex]
Z_{t,p+q+l}(\bmtht, \bm \gamma) = \varepsilon_{t}(\bm\theta)\varepsilon_{t-l}(\bm\theta) - \gamma_l,\quad l = 1, 2, \cdots, m.
  \end{cases}
\end{eqnarray*}
Throughout this paper, we compute $\partial \varepsilon_{t}(\bm\theta) / \partial \bm\theta$ recursively by
\begin{small}
\begin{eqnarray*}
  \frac{\partial \varepsilon_{t}(\bmtht)}{\partial \bmtht} = -
   \tilde X_t - \sum_{j=1}^q \psi_{j} \frac{\partial\varepsilon_{t-j}(\bmtht)}{\partial \bmtht},\quad t = 1, 2, \cdots, n,
\end{eqnarray*}
\end{small}
where $\tilde X_t = ( 1, X_{t-1}, \cdots, X_{t-p}, \varepsilon_{t-1}(\bmtht),   \cdots, \varepsilon_{t-q}(\bmtht))^\top$.

It follows from the Lagrange multiplier technique that
\begin{eqnarray*}
 -2\log L(\bmtht, \bm \gamma) = -2 \sum_{t=m+1}^n \log \{1 + \bm {\lambda}^\top \boldsymbol Z_{t}(\bmtht, \bm \gamma)\},
\end{eqnarray*}
where $\bm {\lambda} =  \bm {\lambda}(\bmtht, \bm \gamma)$ satisfies
\begin{eqnarray*}
\sum_{t=m+1}^n \frac{\boldsymbol Z_{t}(\bmtht, \bm \gamma)}{1+ \bm {\lambda}^\top \boldsymbol Z_{t}(\bmtht, \bm \gamma)} = 0.
\end{eqnarray*}

Since we are interested in testing $\bm\gamma$, we consider the log-profile empirical likelihood function as follows
\begin{eqnarray*}
  \ell(\bm\gamma) = -2\log \{\sup_{\bmtht} L(\bmtht, \bm\gamma)\}.
\end{eqnarray*}
Denote by $\Theta$ the parameters space, which is compact subset of $\mathbb{R}^{p+q+1}$. Suppose the following conditions hold, i.e.,
\begin{itemize}
  \item [\textbf{(C1)}] The true value, say $\bmtht_0$, of $\bmtht$ is an interior point in $\Theta$, and for $\bmtht \in \Theta$, $\phi(z) \neq 0$ and $\psi(z) \neq 0$ when $|z| < 1$, and $\phi(z) = 1 - \sum_{i=1}^p \phi_iz^i$ and $\psi(z) = 1 + \sum_{j=1}^q \psi_jz^j$ have no comment root with $\phi_p \neq 0$ or $\psi_q \neq 0$.

  \item [\textbf{(C2)}] $E(|\varepsilon_t|^{4+\delta}) < \infty$ for some constant $\delta > 0$.

\end{itemize}
Based on the above assumptions, we have the following result.
\begin{theorem}\label{th:001}
  Suppose that $\{\varepsilon_t\}$ is a martingale difference series, i.e. there is no serial correlation existing in $\{\varepsilon_t\}$. Then, under Conditions \textbf{(C1)-(C2)}, we have
  \begin{eqnarray*}
    \ell(0) \overset{d}{\longrightarrow} \chi_{m}^2,\quad \text{as } n \to \infty,
  \end{eqnarray*}
   where `$\overset{d}{\longrightarrow}$' denotes the convergence in distribution, and $\chi_{m}^2$ denotes a chi-squared variable with $m$ degrees of freedom.
\end{theorem}

Based on Theorem \ref{th:001}, we may reject the null hypothesis $\mathcal{H}_0$ if $\ell(0) \ge \chi_{m}^2(1-a)$ at the significance level $a \in (0, 1)$, where $\chi_{m}^2(1-a)$ denotes the ($1-a$)-th quantile of the distribution of $\chi_{m}^2$.

\subsection{Infinite variance innovations}

The ARMA models are usually used in analyzing the daily financial series, which may be heavy tailed. To account for this, we further consider the case in this part that the errors $\varepsilon_t$ follow the GARCH process at \eqref{eqn:errors} with possible infinite variance.

Note that the asymptotical validity of the empirical likelihood-based statistic depends on an assumption that $E(|\varepsilon_t|^{2+\nu}) < \infty$ for some positive $\nu > 0$, which is too strict. Hence, we propose to define the profile weighted empirical likelihood function to account for the infinite variance case for testing $\mathcal{H}_0$ as follows
\begin{eqnarray*}
  \tilde\ell(\bm\gamma) = -2\log \{\sup_{\bmtht} \tilde L(\bmtht, \bm\gamma)\},
\end{eqnarray*}
where
\begin{eqnarray*}
  \tilde L(\bmtht, \bm \gamma) &= \sup{} \Big\{\prod_{t=m+1}^n(Np_t): p_{m+1} \geq 0, \cdots, p_n \geq 0,\\\nonumber
  & \sum_{t=m+1}^n p_t=1,
  \sum_{t=m+1}^n p_{t} \tilde{\boldsymbol Z}_{t}(\bmtht, \bm\gamma)=0\Big\},
\end{eqnarray*}
$\tilde{\bm Z}_{t}(\bmtht, \bm \gamma) = (\tilde Z_{t,1}(\bmtht, \bm\gamma)^\top, \tilde Z_{t,p+q+1}(\bmtht, \bm \gamma), \cdots,\tilde Z_{t,p+q+m}\\(\bmtht, \bm \gamma))^\top$ with
\begin{eqnarray*}
  \begin{cases}
\tilde{\bm Z}_{t, 1}(\bmtht, \bm\gamma) = w_{t-1}^{-2} \varepsilon_{t}(\bm\theta)\frac{\partial \varepsilon_{t}(\bm\theta)}{\partial \bm\theta}, \\[2ex]
\tilde{Z}_{t,p+q+l}(\bmtht, \bm \gamma) = w_{t-1}^{-1}w_{t-1-l}^{-1}\varepsilon_{t}(\bm\theta)\varepsilon_{t-l}(\bm\theta) - \gamma_l,
  \end{cases}
\end{eqnarray*}
for $l = 1, 2, \cdots, m$, and
\begin{eqnarray}\label{eqn:weight}
w_{t} = \max \{M_X, \sum_{i=0}^t e^{-\log^2(i+1)} |X_{t-i}|\}.
\end{eqnarray}
In the sequel we take $M_X$ to be the 90\% sample quantile of $\{|X_{t}|\}$. A similar strategy can be found in  \cite{he2020inference}.

For $\tilde \ell(\bm\gamma)$, replace Condition \textbf{(C2)} with \textbf{(C3)} and further assume \textbf{(C4)} as follows:
\begin{itemize}
  \item [\textbf{(C3)}] $E(w_{t-1}^{-4} \xi_{\rho,t-1}^{4+\delta}) < \infty$ for any $\rho \in (0, 1)$, where $\xi_{\rho, t} = 1 + \sum_{i=1}^\infty \rho^i |X_{t-i}|$ (we suggest to use $\rho = 0.95$ based on simulations), $w_t$ is stationary and $\mathcal{F}_{t}$-measurable, and $\inf_t w_t > 0$. Hereafter, $\delta$ is an arbitrary small positive constant, and $\mathcal{F}_{t}$ denotes the sigma field generated by $\{\eta_s: s \le t\}$, for $t = 1, 2, \cdots, n$.
    \item [\textbf{(C4)}] $\nu^* < 0$, where $\nu^*$ is the Lyapunov exponent of the random matrix $A_{t}$, and
  \begin{eqnarray*}
    \nu^* = \inf \left\{\frac{1}{n} E(\ln\|\mathbf{A}_1\mathbf{A}_2\cdots\mathbf{A}_n\|_{\max}): n = 1, 2, \cdots\right\},
  \end{eqnarray*}
  where $\|\mathbf{A}_1\mathbf{A}_2\cdots\mathbf{A}_n\|_{\max}$ means to maximize the norm of $\mathbf{A}_1\mathbf{A}_2\cdots\mathbf{A}_n$, and
  \begin{eqnarray*}
    &&\mathbf{A}_t =\\
    && \begin{pmatrix}
\tilde a_1^* &b_2 & \ldots &b_{s-1} &\beta_s &a_2&a_3&\ldots &a_r\\
1 & 0 & \ldots  &0 & 0 &0&0& \ldots & 0\\
0 & 1 & \ldots &0 &0 & 0 &0& \ldots & 0\\
\vdots & \vdots & \ddots &\vdots&\vdots & \vdots &\vdots& \ddots &   \vdots\\
0 & 0&\ldots &1 &0 & 0 & 0 & \ldots   & 0\\
\eta_t^2 &0 & \ldots &0 & 0 & 0 &0&\ldots &0 \\
0 &0  & \ldots &0 &0 & 1 &0 & \ldots & 0\\
\vdots &\vdots& \ddots & \vdots &\vdots&\vdots & \vdots & \ddots & \vdots\\
0 &0& \ldots & 0 &0 &0   & \ldots    & 1& 0\\
\end{pmatrix},
\end{eqnarray*}
\end{itemize}
with $\tilde a_1^* = a_1\eta_t^2+b_1$ and $\|\mathbf{A}_t\| = \sup_{|\bm x|=1} |\mathbf{A}_t \bm x|$. We can prove the following result.
\begin{theorem}\label{th:002}
  Suppose that $\{\eta_t\}$ is a sequence of i.i.d. random variables with mean zero and variance one, indicating that there is no serial correlation existing in $\{\varepsilon_t\}$. Then, under Conditions \textbf{(C1)}, \textbf{(C3)}, and \textbf{(C4)}, we have
  \begin{eqnarray*}
    \tilde\ell(0) \overset{d}{\longrightarrow} \chi_{m}^2,\quad \text{as } n \to \infty.
  \end{eqnarray*}
\end{theorem}

\begin{remark}
  Conditions \textbf{(C1)}-\textbf{(C4)} commonly used in the literature. \textbf{(C1)} and \textbf{(C4)} are assumed to guarantee the stationarity of $\{X_t\}$ and $\{\sigma_t\}$, respectively; see, e.g., \cite{ling2007self} and \cite{ma2021test}. \textbf{(C3)} allows the weight to reduce the moment effect of $\sigma_t$. By \cite{ma2021test}, we have that the weigh defined in \eqref{eqn:weight} satisfies Condition \textbf{(C3)}. Under \textbf{(C3)}, although $\sigma_t$ may have infinite variance, the result of Theorem 2 still holds, fortunately.
\end{remark}

\begin{remark}
  By `infinite variance' we mean that $E(\varepsilon_t^2|\mathcal{F}_{t-1})$ tends to infinite almost surely as $t\to \infty$, noting that $E(\varepsilon_t^2|\mathcal{F}_{t-1}) = \sigma_t^2$, while Theorem \ref{th:001} depends on Condition \textbf{(C2)}, i.e., $E(|\varepsilon_t|^{4+\delta}) < \infty$ for some constant $\delta > 0$. Hereafter, $\mathcal{F}_{t}$ denotes the sigma field generated by $\{\eta_s: s \le t\}$.
\end{remark}

\begin{remark}
Compared with the unweighted empirical likelihood test, which requires at least finite 4th order moment on the data process $\{X_t\}$, the weighted empirical likelihood test needs no moment condition on $\{X_t\}$, but instead the condition $E(w_{t-1}^{-4} \xi_{\rho,t-1}^{4+\delta}) < \infty$ to guarantee the chi-squared limit distribution as indicated in Theorem \ref{th:002}.
\end{remark}

Theorem \ref{th:002} indicates that through controlling the effect of the error variance, the weighted log-empirical likelihood ratio still has a standard limit distribution. Based on Theorem \ref{th:002}, we may similarly reject the null hypothesis $\mathcal{H}_0$ if $ \tilde\ell(0) \ge \chi_{m}^2(1-a)$ at the significance level $a \in (0, 1)$.

\vskip 0.1 in
\section{Simulation results}
\label{Simulation}

In this section, we carry out some simulation experiments to illustrate the finite sample properties of the proposed empirical likelihoods when the variance of $\varepsilon _{t}$ is finite or infinite. For the sake of comparison, we also report the result of the $\tilde Q$ statistic in \cite{zhu2016bootstrapping}.

The simulated data $\{X_{t}\}_{t=1}^{n}$ are generated from:
\begin{equation*}
\begin{cases}
X_{t}=\mu + \phi X_{t-1} +\psi \varepsilon _{t-1} + \varepsilon _{t}, \\[2ex]
\varepsilon_{t}=\eta _{t}\sigma _{t},~\sigma _{t}^{2}=\omega +a\varepsilon
_{t-1}^{2}+b\sigma _{t-1}^{2},
\end{cases}%
\end{equation*}%
where $\varepsilon _{t}$ follows a GARCH(1,1) process, $\eta_{t}= (\frac{c}{\sqrt{n}} e_{t-1} + e_t) / \sqrt{1+ (\frac{c}{\sqrt{n}})^2}$,
and $\{e_t\}$ is a sequence of $i.i.d.$ random variables generated from the standard normal distribution. $c$ is taken from $\{0,~5,~10,~15\}$ with $c=0$ standing for the validity of $\mathcal{H}_0$, while $c=5$ or $c=10$ or $c=15$ representing that the local alternative hypothesis of $\mathcal{H}_0$ holds. We set $\phi = 0.3$, $\psi = 0.4$, $\omega=0.2$, and consider two different intercepts $\mu$, i.e., 0, 0.5. For the GARCH process of $\varepsilon _{t}$, we choose $(a,~b)=(0.1,~0.15)$ to represent the variance of $\varepsilon _{t}$ being finite, while $(a,~b)=(0.33,~0.66)$ to imply the infinite variance of $\varepsilon _{t}$ approximately. Note that when $a+b$ is close to 1, we have $\sigma _{t}^2 \to \infty$ as $t \to \infty$.

For simplicity, we by `EL' mean the naive empirical likelihood method, by `WeL' the weighted empirical likelihood method, and by $\tilde Q$ the random weighted bootstrapping statistic given in \cite{zhu2016bootstrapping}, respectively. We investigate the performance of $\tilde Q$, EL, and WeL in testing whether the residuals are correlated at lags $m = 2$ or $m = 6$. Note that the diagonal matrix $W^*$ for $\tilde Q$ is taken to be the identity matrix of order $m$, and the random weights are generated from the exponential distribution with parameter 1 ensuring that the weights have means one and variances one. The other settings for the random weighted bootstrap are the same as those in  \cite{zhu2016bootstrapping}.

Tables 3.1-3.4 report the empirical ratios of rejecting $\mathcal{H}_0$ based on 2000 replications at significance levels $\tau =0.1$ and $0.05$. Three sample sizes, i.e., $n =400,800$, and $1200$, are considered, and there are four findings. (i) For the case of $(a,b)=(0.1,0.15)$, the sizes of both EL and WeL are very close to the nominal levels, noting that EL is better than WeL. (ii) For the case of $(a,b)=(0.33,0.66)$, as expected, WeL performs the best, but is slightly over-sized. Fortunately, its size decreases as $n$ increases. Note that EL is highly over-sized and its size seems not to be convergent as the sample size increases. (iii) There is a size distortion for $\tilde Q$ in our reported cases compared to the proposed empirical likelihood methods. (iv) Both EL and WeL have nontrivial local powers, and their powers increase as the value of $c$ increases.

It is noted that WeL suffers from a loss of power owing to the usage of the weighting technique compared to EL. Both EL and WeL are sightly over-sized for the finite variance case when $m = 6$, which indicates that the empirical likelihood-based testing methods, i.e., EL and WeL, are affected by the dimension of the auxiliary vectors. Similar phenomena have been observed in the literature. In practice, one may increase the precision of the chi-square approximation through adding proper pseudo-observations; see, e.g., \citep{chen2008adjusted} and \cite{liu2010adjusted} for details.

\begin{table*}
 \setlength{\belowcaptionskip}{.2cm}
 {\bf TABLE 3.1}\\
\centering The finite variance case with $(a,b)=(0.1,0.15)$, $(\phi,\psi)=(0.3,0.4)$ and $m=2$.
\centering \renewcommand{\arraystretch}{0.8} 
\setlength{\tabcolsep}{4mm}{\
\begin{tabular}{p{0.8cm}ccp{0.8cm}p{0.8cm}p{0.8cm}cp{0.8cm}p{0.8cm}p{0.8cm}}
\toprule \multirow{2}{*}{$\mu$} & \multirow{2}{*}{$n$} & %
\multirow{2}{*}{$c$} & \multicolumn{3}{c}{$\tau =0.1$} &  &
\multicolumn{3}{c}{$\tau =0.05$} \\ \cline{4-10}\\[-1.5ex]
    &  &   & $\tilde{Q}$ & EL & WeL & & $\tilde{Q}$ & EL & WeL \\
  \midrule
  \multirow{16}{*}{0}
    & \multirow{4}{*}{400}
         &        0                  & 0.017 & 0.094 & 0.104 && 0.006 & 0.049 & 0.053\\
         &&       5                  & 0.151 & 0.227 & 0.140 && 0.083 & 0.141 & 0.078\\
         &&       10                 & 0.616 & 0.607 & 0.314 && 0.502 & 0.488 & 0.213\\
         &&       15                 & 0.954 & 0.936 & 0.717 && 0.930 & 0.883 & 0.593\\
       \cline{2-10}
    &  \multirow{4}{*}{800}
         &        0                  & 0.027 & 0.104 & 0.108 && 0.012 & 0.055 & 0.057\\
         &&       5                  & 0.134 & 0.198 & 0.120 && 0.083 & 0.124 & 0.069\\
         &&       10                 & 0.517 & 0.541 & 0.226 && 0.405 & 0.408 & 0.133\\
         &&       15                 & 0.918 & 0.911 & 0.532 && 0.873 & 0.844 & 0.398\\
       \cline{2-10}
    &  \multirow{4}{*}{1200}
         &        0                  & 0.028 & 0.091 & 0.102 && 0.013 & 0.047 & 0.054\\
         &&       5                  & 0.118 & 0.177 & 0.126 && 0.068 & 0.101 & 0.062\\
         &&       10                 & 0.464 & 0.510 & 0.219 && 0.364 & 0.374 & 0.135\\
         &&       15                 & 0.883 & 0.871 & 0.459 && 0.812 & 0.791 & 0.341\\
     \hline
   \multirow{16}{*}{0.5}
    & \multirow{4}{*}{400}
         &        0                  & 0.017  & 0.103  & 0.099  &  & 0.005  & 0.052  & 0.048 \\
         &&       5                  & 0.131  & 0.215  & 0.126  &  & 0.073  & 0.125  & 0.072  \\
         &&       10                 & 0.597  & 0.597  & 0.374  &  & 0.496  & 0.473  & 0.255   \\
         &&       15                 & 0.954  & 0.936  & 0.792  &  & 0.921  & 0.877  & 0.686  \\
      \cline{2-10}
    & \multirow{4}{*}{800}
         &        0                  & 0.027  & 0.108  & 0.106  &  & 0.013  & 0.056  & 0.055  \\
         &&       5                  & 0.124  & 0.202  & 0.124  &  & 0.069  & 0.121  & 0.064  \\
         &&       10                 & 0.504  & 0.529  & 0.270  &  & 0.394  & 0.399  & 0.183  \\
         &&       15                 & 0.902  & 0.898  & 0.649  &  & 0.849  & 0.830  & 0.524 \\
      \cline{2-10}
     & \multirow{4}{*}{1200}
         &        0                  & 0.024 & 0.092 & 0.095 && 0.014 & 0.047 & 0.042\\
         &&       5                  & 0.113 & 0.178 & 0.111 && 0.065 & 0.102 & 0.053\\
         &&       10                 & 0.465 & 0.512 & 0.222 && 0.358 & 0.374 & 0.141\\
         &&       15                 & 0.881 & 0.875 & 0.566 && 0.813 & 0.790 & 0.434\\
  \bottomrule
 \end{tabular}}
\end{table*}

\begin{table*}
 \setlength{\belowcaptionskip}{.2cm}
 {\bf TABLE 3.2}\\
\centering The finite variance case with $(a, b) = (0.1, 0.15)$, $(\phi,\psi)=(0.3,0.4)$ and $m=6$.
\centering \renewcommand{\arraystretch}{0.8} 
\setlength{\tabcolsep}{4mm}{\
\begin{tabular}{p{0.8cm}ccp{0.8cm}p{0.8cm}p{0.8cm}cp{0.8cm}p{0.8cm}p{0.8cm}}
\toprule \multirow{2}{*}{$\mu$} & \multirow{2}{*}{$n$} & %
\multirow{2}{*}{$c$} & \multicolumn{3}{c}{$\tau =0.1$} &  &
\multicolumn{3}{c}{$\tau =0.05$} \\ \cline{4-10} \\[-1.5ex]
    &  &   & $\tilde{Q}$ & EL & WeL & & $\tilde{Q}$ & EL & WeL \\
  \midrule
  \multirow{16}{*}{0}
    & \multirow{4}{*}{400}
         &        0                  & 0.000 & 0.104 & 0.128 && 0.000 & 0.054 & 0.073\\
         &&       5                   & 0.000 & 0.184 & 0.178 && 0.000 & 0.115 & 0.106\\
         &&       10                  & 0.004 & 0.484 & 0.329 && 0.001 & 0.351 & 0.220\\
         &&       15                  & 0.033 & 0.855 & 0.617 && 0.007 & 0.766 & 0.489\\
       \cline{2-10}
    &  \multirow{4}{*}{800}
         &        0                  & 0.000 & 0.110 & 0.127 && 0.000 & 0.054 & 0.069\\
         &&       5                  & 0.000 & 0.170 & 0.144 && 0.000 & 0.095 & 0.091\\
         &&       10                 & 0.002 & 0.418 & 0.261 && 0.000 & 0.288 & 0.171\\
         &&       15                 & 0.029 & 0.793 & 0.519 && 0.005 & 0.691 & 0.388\\
       \cline{2-10}
    &  \multirow{4}{*}{1200}
         &        0                  & 0.000 & 0.097 & 0.107 && 0.000 & 0.049 & 0.058\\
         &&       5                  & 0.002 & 0.140 & 0.135 && 0.000 & 0.070 & 0.071\\
         &&       10                 & 0.003 & 0.364 & 0.234 && 0.001 & 0.244 & 0.138\\
         &&       15                 & 0.021 & 0.742 & 0.486 && 0.003 & 0.624 & 0.353\\
     \hline
     \multirow{16}{*}{0.5}
    & \multirow{4}{*}{400}
         &        0                & 0.000 & 0.125 & 0.139 && 0.000 & 0.073 & 0.081\\
         &&       5                & 0.000 & 0.177 & 0.177 && 0.000 & 0.109 & 0.110\\
         &&       10               & 0.004 & 0.468 & 0.378 && 0.001 & 0.349 & 0.253\\
         &&       15               & 0.032 & 0.845 & 0.691 && 0.007 & 0.763 & 0.566\\
      \cline{2-10}
    & \multirow{4}{*}{800}
         &        0                & 0.000 & 0.117 & 0.126 && 0.000 & 0.061 & 0.065\\
         &&       5                & 0.000 & 0.164 & 0.156 && 0.000 & 0.091 & 0.088\\
         &&       10               & 0.002 & 0.412 & 0.290 && 0.000 & 0.286 & 0.190\\
         &&       15               & 0.029 & 0.790 & 0.603 && 0.005 & 0.687 & 0.472\\
      \cline{2-10}
    & \multirow{4}{*}{1200}
         &        0                & 0.000 & 0.096 & 0.107 && 0.000 & 0.050 & 0.055\\
         &&       5                & 0.001 & 0.138 & 0.137 && 0.000 & 0.071 & 0.070\\
         &&       10               & 0.003 & 0.364 & 0.273 && 0.001 & 0.241 & 0.178\\
         &&       15               & 0.020 & 0.747 & 0.558 && 0.003 & 0.622 & 0.430\\
  \bottomrule
 \end{tabular}}
\end{table*}

\begin{table*}
 \setlength{\belowcaptionskip}{.2cm}
 {\bf TABLE 3.3}\\
\centering The infinite variance case with $(a, b)=(0.33,0.66),~(\phi, \psi) = (0.3, 0.4)$ and $m=2$.
\centering \renewcommand{\arraystretch}{0.8} 
\setlength{\tabcolsep}{4mm}{\
\begin{tabular}{p{0.8cm}ccp{0.8cm}p{0.8cm}p{0.8cm}cp{0.8cm}p{0.8cm}p{0.8cm}}
\toprule \multirow{2}{*}{$\mu$} & \multirow{2}{*}{$n$} & %
\multirow{2}{*}{$c$} & \multicolumn{3}{c}{$\tau =0.1$} &  &
\multicolumn{3}{c}{$\tau =0.05$} \\ \cline{4-10}\\[-1.5ex]
    &  &   & $\tilde{Q}$ & EL & WeL & & $\tilde{Q}$ & EL & WeL \\
  \midrule
  \multirow{16}{*}{0}
    & \multirow{4}{*}{400}
         &        0                  & 0.012  & 0.211  & 0.106  &  & 0.006  & 0.148  & 0.057  \\
         &&       5                  & 0.075  & 0.299  & 0.141  &  & 0.035  & 0.213  & 0.077  \\
         &&       10                 & 0.295  & 0.471  & 0.335  &  & 0.206  & 0.359  & 0.233  \\
         &&       15                 & 0.583  & 0.658  & 0.633  &  & 0.479  & 0.522  & 0.544  \\
       \cline{2-10}
    &  \multirow{4}{*}{800}
         &        0                  & 0.020  & 0.294  & 0.114  &  & 0.013  & 0.234  & 0.064  \\
         &&       5                  & 0.059  & 0.354  & 0.138  &  & 0.033  & 0.268  & 0.077  \\
         &&       10                 & 0.201  & 0.487  & 0.257  &  & 0.133  & 0.385  & 0.185  \\
         &&       15                 & 0.445  & 0.596  & 0.518  &  & 0.343  & 0.493  & 0.394  \\
       \cline{2-10}
    &  \multirow{4}{*}{1200}
         &        0                  & 0.014 & 0.360 & 0.115 && 0.003 & 0.292 & 0.066\\
         &&       5                  & 0.043 & 0.401 & 0.144 && 0.020 & 0.325 & 0.080\\
         &&       10                 & 0.141 & 0.477 & 0.257 && 0.084 & 0.389 & 0.181\\
         &&       15                 & 0.328 & 0.588 & 0.487 && 0.234 & 0.480 & 0.377\\
     \hline
   \multirow{16}{*}{0.5}
    & \multirow{4}{*}{400}
         &        0                  & 0.021  & 0.217  & 0.110  &  & 0.010  & 0.148  & 0.064  \\
         &&       5                  & 0.069  & 0.283  & 0.149  &  & 0.039  & 0.213  & 0.082 \\
         &&       10                 & 0.290  & 0.478  & 0.351  &  & 0.192  & 0.371  & 0.246 \\
         &&       15                 & 0.591  & 0.660  & 0.632  &  & 0.478  & 0.557  & 0.540  \\
      \cline{2-10}
    & \multirow{4}{*}{800}
         &        0                  & 0.016  & 0.298  & 0.111  &  & 0.009  & 0.229  & 0.064  \\
         &&       5                  & 0.057  & 0.352  & 0.130  &  & 0.028  & 0.287  & 0.081   \\
         &&       10                 & 0.201  & 0.465  & 0.285  &  & 0.133  & 0.369  & 0.187   \\
         &&       15                 & 0.454  & 0.599  & 0.540  &  & 0.354  & 0.484  & 0.424   \\
      \cline{2-10}
    & \multirow{4}{*}{1200}
         &        0                  & 0.014 & 0.363 & 0.107 && 0.003 & 0.299 & 0.062\\
         &&       5                  & 0.043 & 0.400 & 0.141 && 0.021 & 0.325 & 0.081\\
         &&       10                 & 0.141 & 0.487 & 0.265 && 0.084 & 0.399 & 0.191\\
         &&       15                 & 0.328 & 0.601 & 0.485 && 0.233 & 0.496 & 0.378\\
  \bottomrule
 \end{tabular}}
\end{table*}

\begin{table*}
 \setlength{\belowcaptionskip}{.2cm}
 {\bf TABLE 3.4}\\
\centering The infinite variance case with $(a, b)=(0.33,0.66),~(\phi,\psi)=(0.3,0.4)$ and $m=6$.
\centering \renewcommand{\arraystretch}{0.8} 
\setlength{\tabcolsep}{4mm}{\
\begin{tabular}{p{0.8cm}ccp{0.8cm}p{0.8cm}p{0.8cm}cp{0.8cm}p{0.8cm}p{0.8cm}}
\toprule \multirow{2}{*}{$\mu$} & \multirow{2}{*}{$n$} & %
\multirow{2}{*}{$c$} & \multicolumn{3}{c}{$\tau =0.1$} &  &
\multicolumn{3}{c}{$\tau =0.05$} \\ \cline{4-10}\\[-1.5ex]
    &  &   & $\tilde{Q}$ & EL & WeL & & $\tilde{Q}$ & EL & WeL \\
  \midrule
  \multirow{16}{*}{0}
    & \multirow{4}{*}{400}
         &        0      & 0.000 & 0.202 & 0.136 && 0.000 & 0.133 & 0.079\\
         &&       5      & 0.000 & 0.291 & 0.185 && 0.000 & 0.209 & 0.111\\
         &&       10     & 0.000 & 0.436 & 0.374 && 0.000 & 0.338 & 0.261\\
         &&       15     & 0.003 & 0.608 & 0.560 && 0.000 & 0.504 & 0.471\\
       \cline{2-10}
    &  \multirow{4}{*}{800}
         &        0      & 0.000 & 0.255 & 0.127 && 0.000 & 0.183 & 0.074\\
         &&       5      & 0.000 & 0.314 & 0.181 && 0.000 & 0.232 & 0.111\\
         &&       10     & 0.000 & 0.436 & 0.338 && 0.000 & 0.345 & 0.238\\
         &&       15     & 0.003 & 0.556 & 0.559 && 0.000 & 0.472 & 0.453\\
       \cline{2-10}
    &  \multirow{4}{*}{1200}
         &        0      & 0.000 & 0.292 & 0.122 && 0.000 & 0.223 & 0.073\\
         &&       5      & 0.000 & 0.346 & 0.188 && 0.000 & 0.262 & 0.115\\
         &&       10     & 0.001 & 0.416 & 0.323 && 0.000 & 0.330 & 0.235\\
         &&       15     & 0.001 & 0.513 & 0.522 && 0.000 & 0.422 & 0.402\\
     \hline
   \multirow{16}{*}{0.5}
    & \multirow{4}{*}{400}
         &        0     & 0.000 & 0.204 & 0.130 && 0.000 & 0.136 & 0.077\\
         &&       5     & 0.000 & 0.278 & 0.196 && 0.000 & 0.200 & 0.112\\
         &&       10    & 0.000 & 0.436 & 0.383 && 0.000 & 0.326 & 0.281\\
         &&       15    & 0.003 & 0.604 & 0.565 && 0.000 & 0.495 & 0.491\\
      \cline{2-10}
    & \multirow{4}{*}{800}
         &        0     & 0.000     & 0.262 & 0.120 && 0.000 & 0.188 & 0.069\\
         &&       5     & 0.000     & 0.312 & 0.174 && 0.000 & 0.233 & 0.110\\
         &&       10    & 0.000     & 0.432 & 0.330 && 0.000 & 0.345 & 0.231\\
         &&       15    & 0.003 & 0.550 & 0.531 && 0.000 & 0.469 & 0.428\\
               \cline{2-10}
    & \multirow{4}{*}{1200}
         &        0     & 0.000 & 0.312 & 0.115 && 0.000 & 0.238 & 0.066\\
         &&       5     & 0.000 & 0.352 & 0.190 && 0.000 & 0.268 & 0.124\\
         &&       10    & 0.000 & 0.426 & 0.331 && 0.000 & 0.337 & 0.236\\
         &&       15    & 0.000 & 0.525 & 0.514 && 0.000 & 0.425 & 0.405\\
  \bottomrule
 \end{tabular}}
\end{table*}

\section{Two applications}
\label{applications}

In this section, we conduct two real analyses based on modelling the monthly exchange rate on the stock market and the daily PM2.5 data in different cities by using the ARMA model discussed in this paper.

\subsection{The exchange rate on the stock market}

We first collected the monthly exchange rate of eight countries including emerging and developed countries. The currencies of emerging countries that we use are: Indian rupee (INR), Malaysian ringgit (MYR), South Korea Won (KRW) and Thai baht
(THB); the currencies in developed countries include: Canadian dollar (CAD), British sterling (GBP), Euro (EUR) and Japanese yen (JPY). The stock indices are: S$\&$P/TSX (Canada), DAX (Germany), Nifty 50 (India), Nikkei 225 (Japan), FTSE KLCI (Malaysia), KOSPI Composite Index (South Korea), SET 50 (Thailand) and FTSE 100 (UK). All data are downloaded from \emph{investing.com} and \emph{Yahoo Finance}. We then transform all data by using $\log (\frac{P_{t}}{P_{t-1}})$, where $P_{t}$ is the exchange rate at time $t$, so $X_{t}$ represent the exchange rate return in our model.

\begin{table*}
\setlength{\belowcaptionskip}{.2cm}  \textbf{TABLE 4.1}\\
 \centering The $p$-values of different tests with the monthly the stock market data, where EL(2) stands for the EL method with $m = 2$, and EL(6) is for the EL method with $m = 6$.
\renewcommand{\arraystretch}{0.8} 
\begin{threeparttable}
 \setlength{\tabcolsep}{1.6mm}{
 \begin{tabular}{crccclllllccc}
  \toprule
   \quad          & Country &Time                 &  $\tilde Q(2)$ &  $\tilde Q(6)$  & EL(2) & EL(6)     &  WeL(2) & WeL(6)   \\
    \hline
    \quad         & India   & $1996.01-2020.04$   &0.8111&0.7866&$0.0759^{\ast}$&0.3718&0.5537&$0.0014^{\ast\ast\ast}$\\
    \quad         & Malaysia& $2002.07-2020.04$   &0.8080&0.8312&$0.0008^{\ast\ast\ast}$&$0.0000^{\ast\ast\ast}$&0.8823&0.1495\\
    \quad         & Korea   & $1997.08-2020.04$    &0.5966&0.7337&$0.0813^{\ast}$&$0.0000^{\ast\ast\ast}$&0.5556&$0.0191^{\ast\ast}$\\
    \quad         & Thailand& $2003.10-2020.04$   &0.7602&0.7953&0.3750&$0.0681^{\ast}$&1.0000&0.2412\\
    \quad         & Canada  & $1990.02-2020.04$   &0.7660&0.6682&0.1523&$0.0227^{\ast\ast}$&0.3697&0.3164\\
    \quad         & UK      & $2001.03-2020.04$   &0.4467&0.5571&0.2771&$0.0140^{\ast\ast}$&0.6344&$0.0812^{\ast}$\\
    \quad         & Germany & $1990.02-2020.04$   &0.6805&0.6517&0.6493&$0.0378^{\ast\ast}$&0.4956&0.1433\\
    \quad         & Japan   & $1995.02-2020.04$   &0.3849&0.7115&$0.0003^{\ast\ast\ast}$&$0.0002^{\ast\ast\ast}$&0.7796&0.1050\\
    \bottomrule
    \end{tabular}}
     \begin{tablenotes}
        \footnotesize
        \item Significance levels: $ ^{\ast} p < 0.1, ^{\ast \ast} p < 0.05, ^{\ast \ast \ast} p< 0.01.$
      \end{tablenotes}
  \end{threeparttable}
\end{table*}

\begin{figure}[h]
\flushleft
\includegraphics[scale = 0.3]{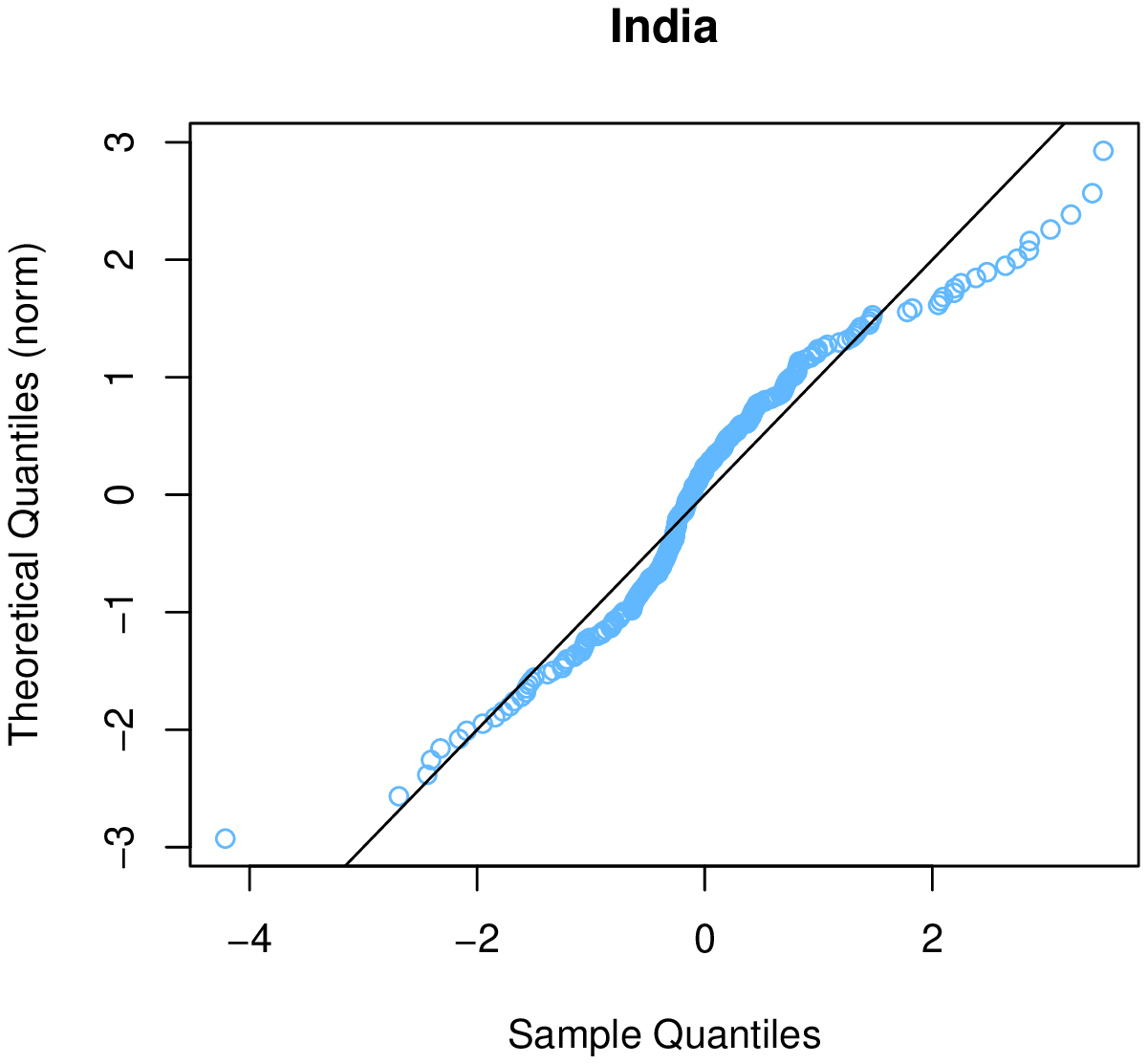}
\includegraphics[scale = 0.3]{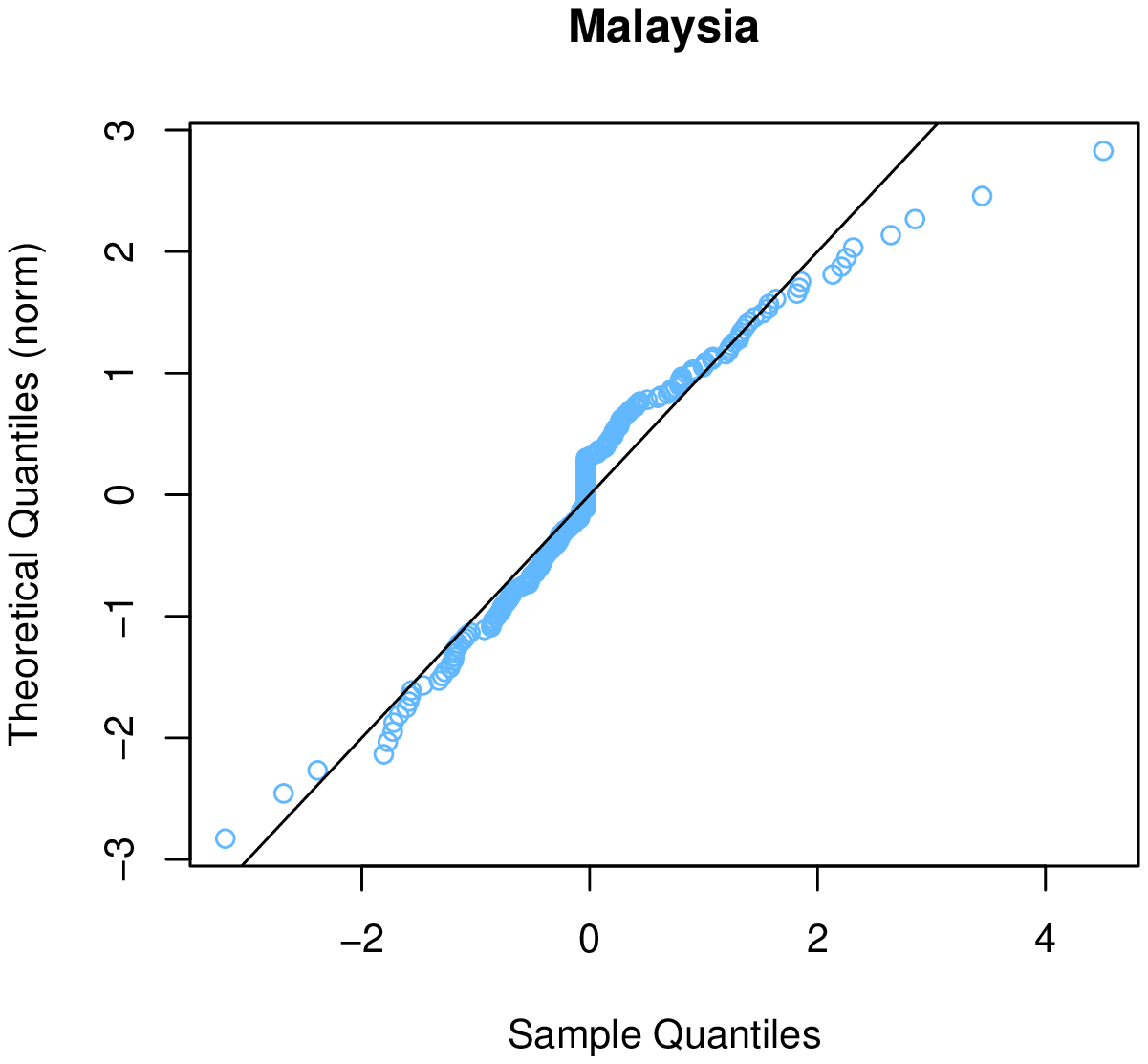}
\includegraphics[scale = 0.3]{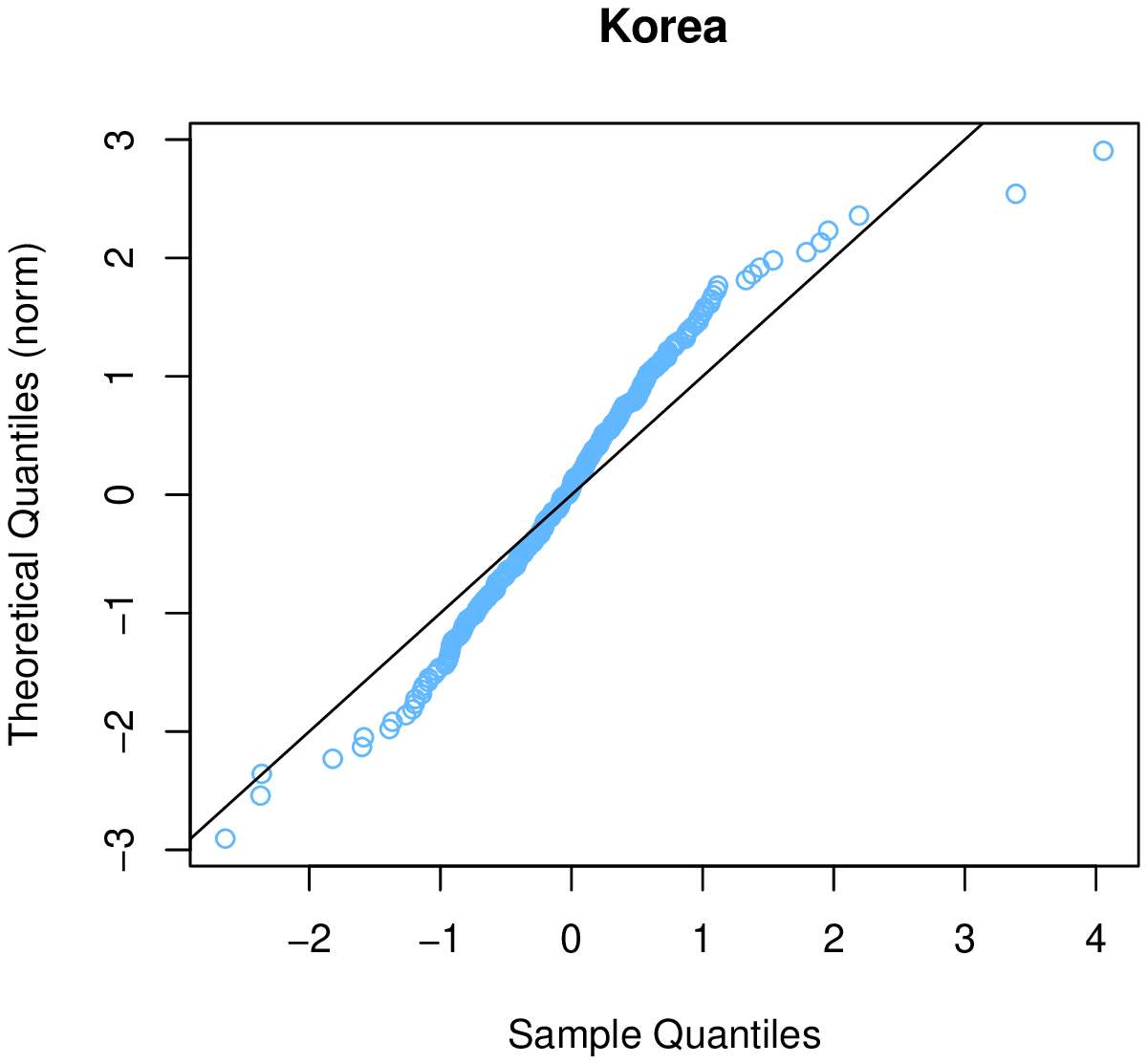}
\includegraphics[scale = 0.3]{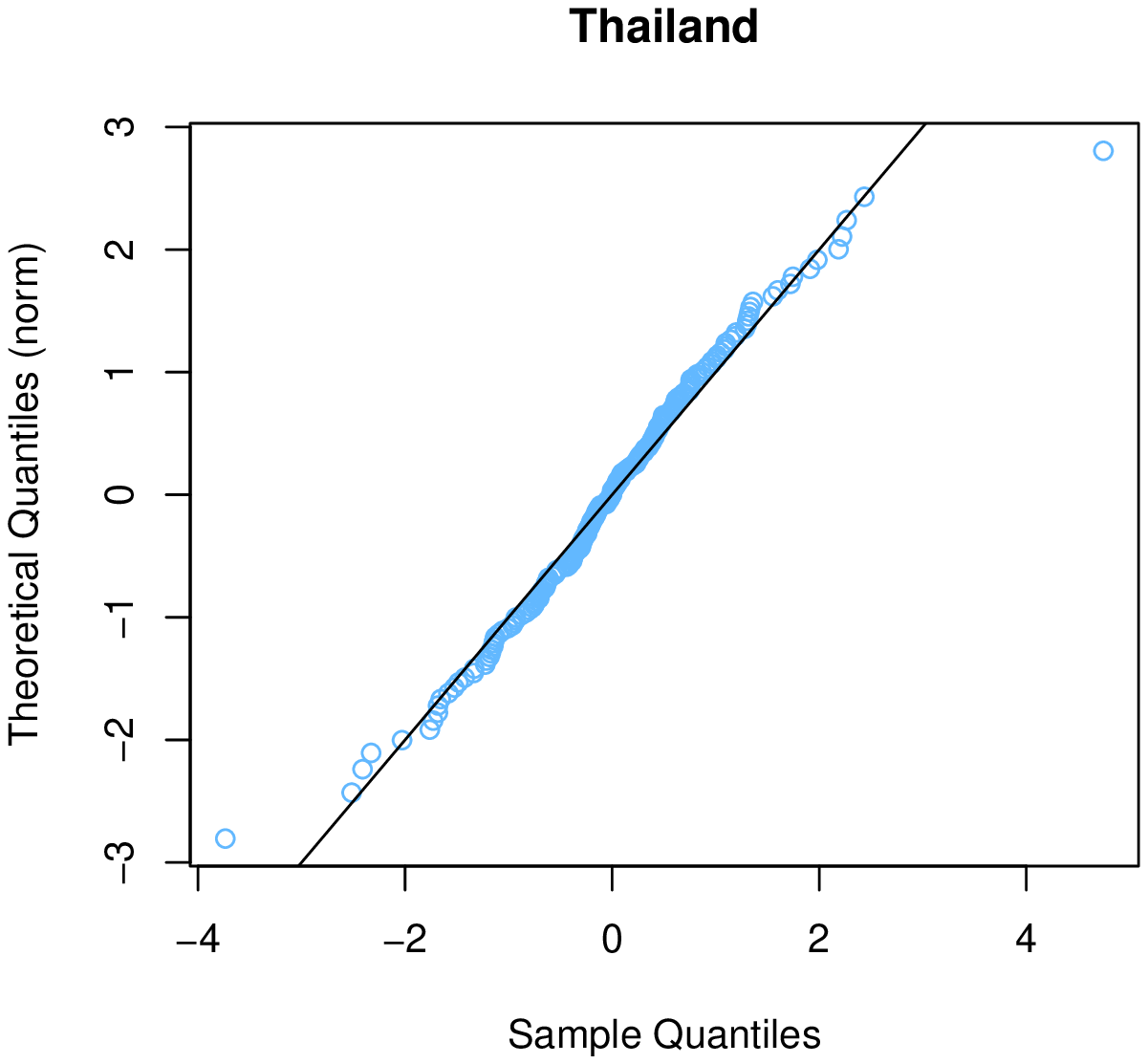}
\includegraphics[scale = 0.3]{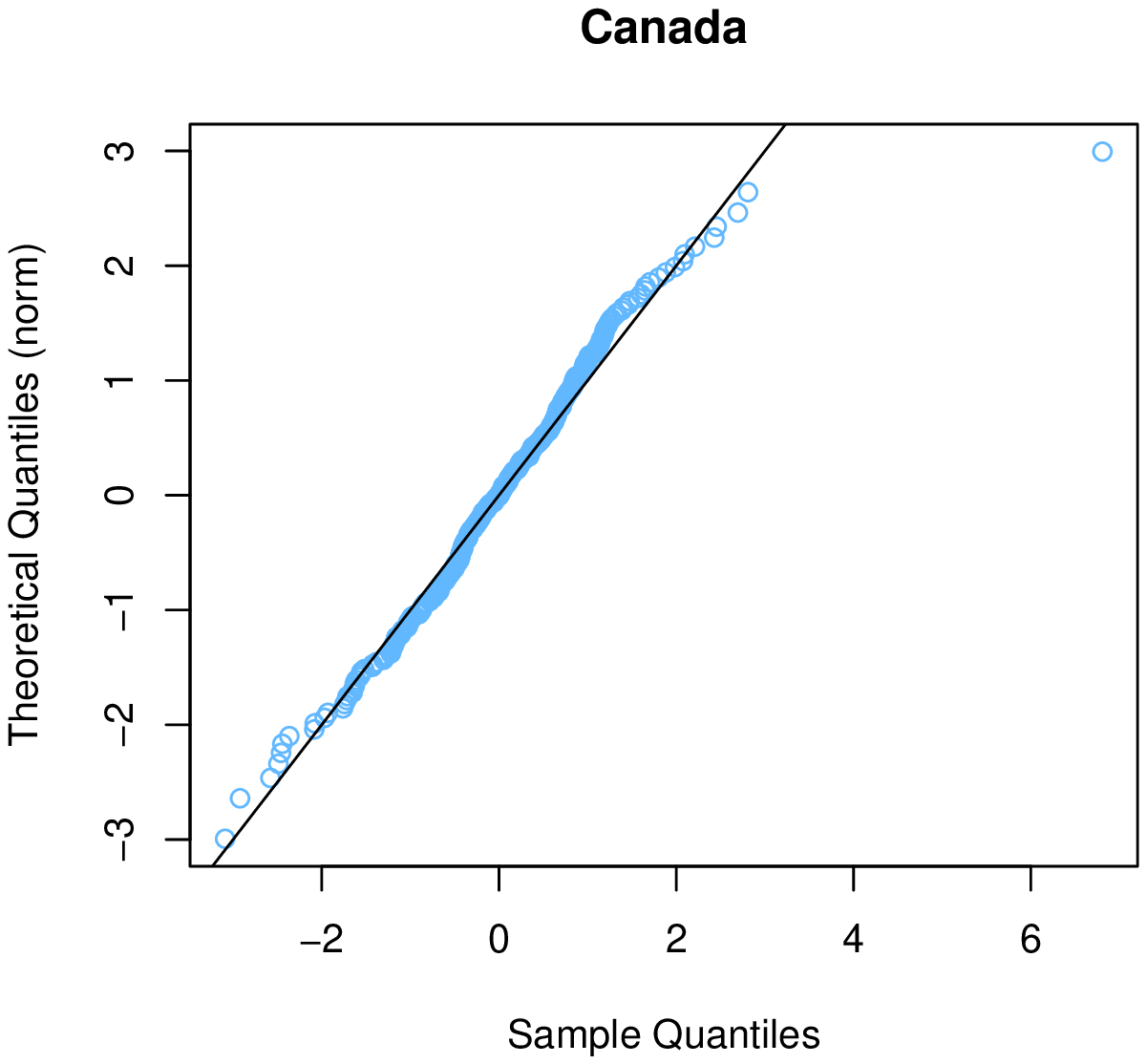}
\includegraphics[scale = 0.3]{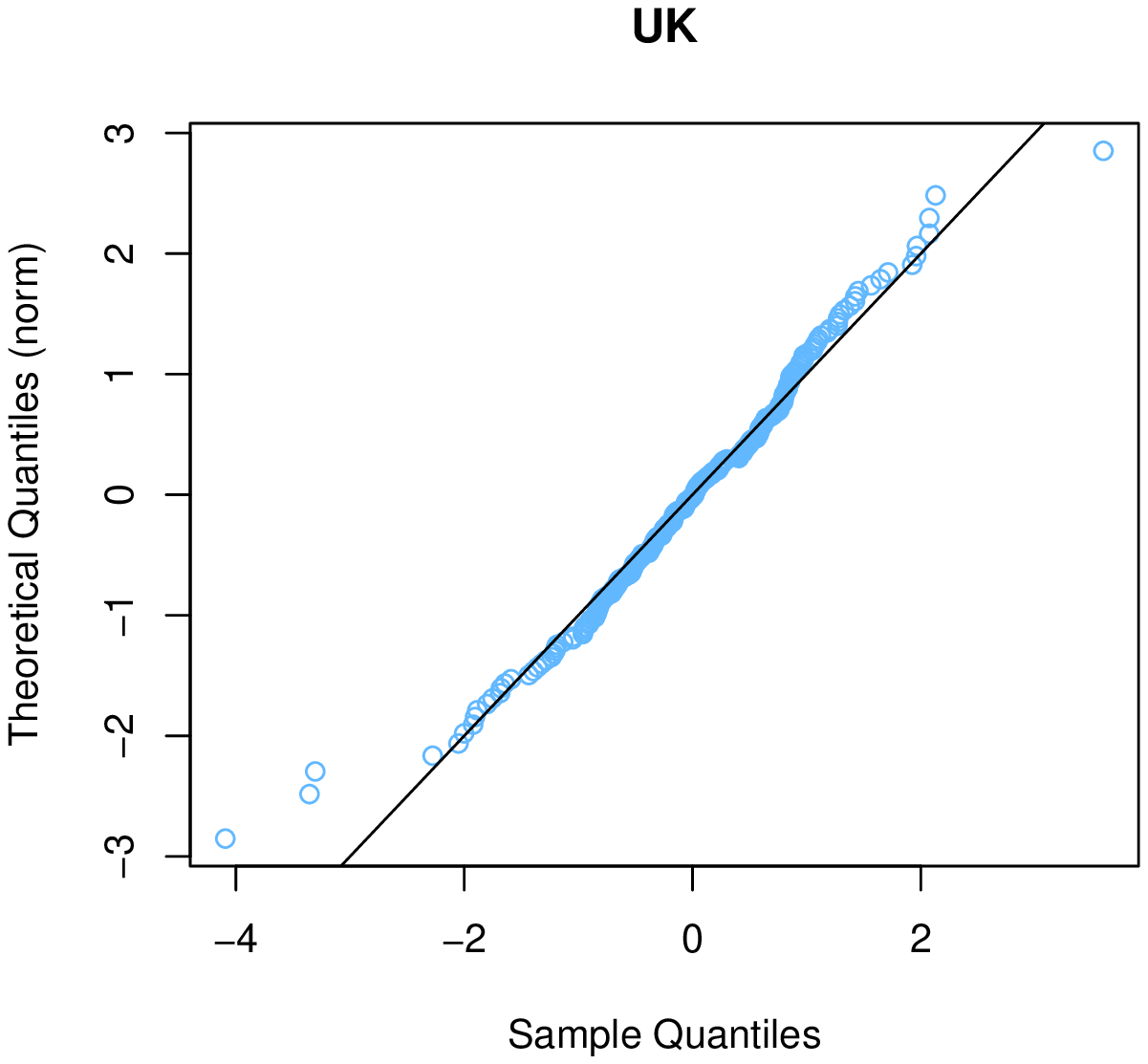}
\includegraphics[scale = 0.3]{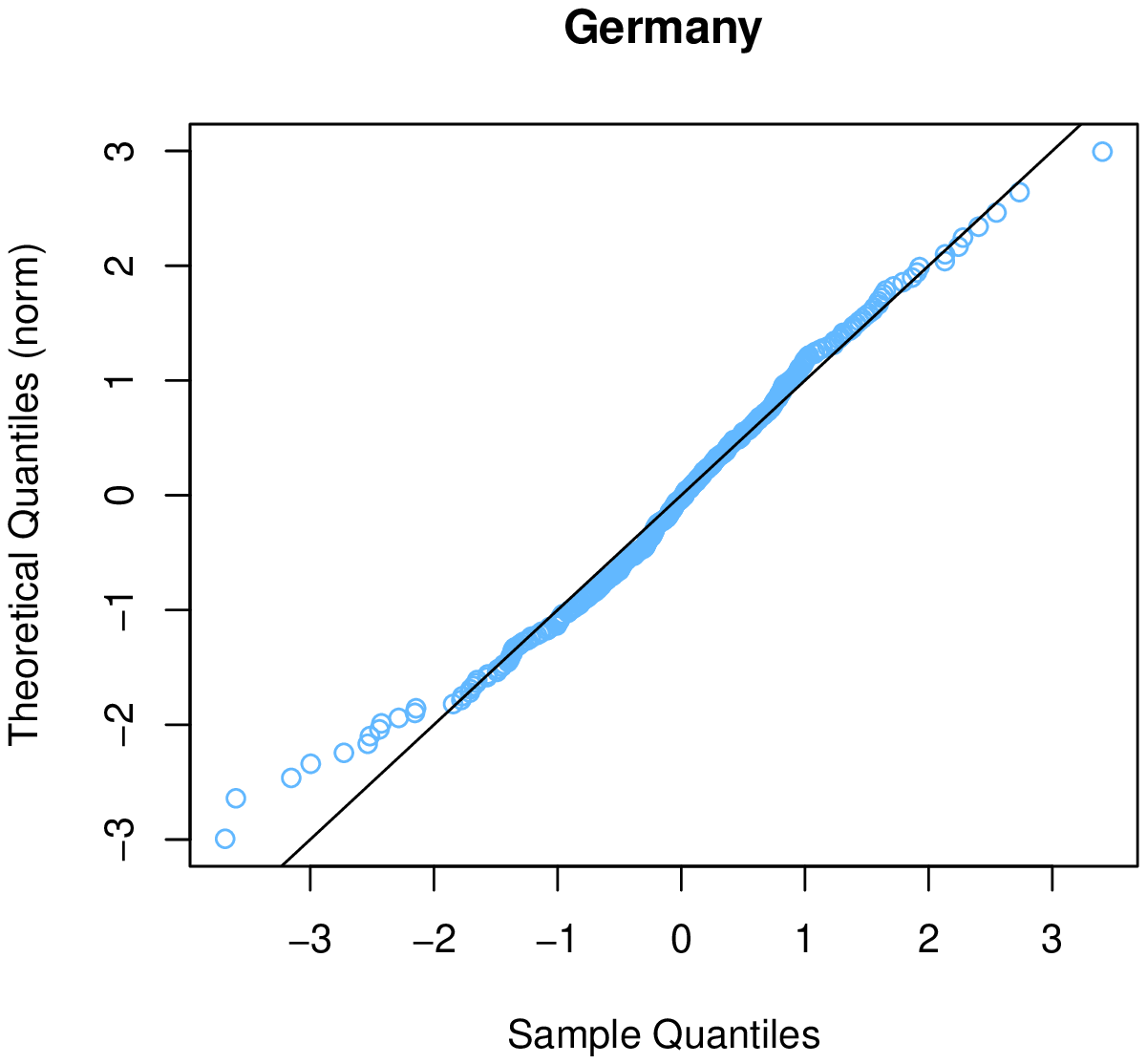}
\includegraphics[scale = 0.3]{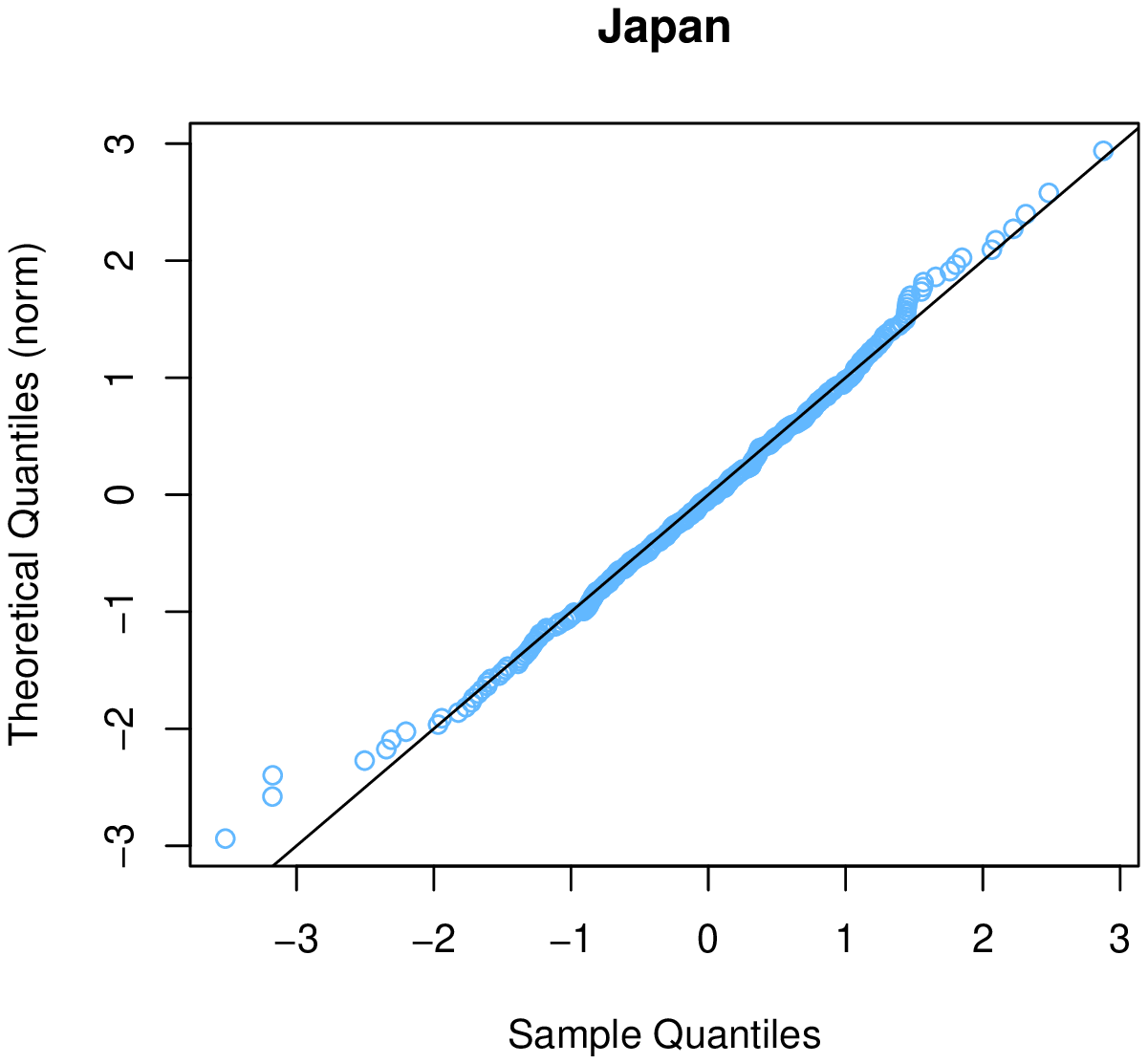}
\caption{QQ-plots for residuals of the monthly exchange rate data from eight countries.}
\label{Fig:Auto1}
\end{figure}

The time spans of the data sets of these eight countries are summarized in Table 4.1. We check the ARCH effect of these data by using the Lagrange multiplier procedure suggested in \cite{engle1982autoregressive}, and found that the $p$-values are 0.0013, 0.0000, 0.0000, 0.0004, 0.027, 0.0000, 0.0249, 0.0117 for the monthly exchange rates of India, Malaysia, Korea, Thailand, Canada, UK, Germany, Japan, respectively. This shows the rationality of fitting these data by using the GARCH-type errors.

To ensure that we use the appropriate test, it is important to check if there is any heavy tail in residuals. In fact, as pointed out in \cite{ibragimov2013emerging}, the heavy-tail feature is of key interest to risk managers, financial regulators, financial stability analysts and policy makers. Several recent studies have suggested that many financial variables may be driven by infinite-variance innovations. For example, studies by \cite{Mandelbrot1963}, \cite{boothe1987statistical}, \cite{koedijk1992tail}, \cite{akgiray1988distribution}, \cite{falk2003testing}, \cite{ibragimov2013emerging} provide evidence for infinite variance behavior in exchange rate return. We show their QQ-plots in Figure \ref{Fig:Auto1} with the standard normal distribution being compared. It seems that the distributions of these monthly data likely do \emph{not} have infinite variances. 

We fit the real data by using \emph{auto.arima.R} contained in the R package `forecast', and then test the possibility of existing serial correlation in the estimated residuals. All results of $\tilde Q$, EL and WeL are summarized in Table 4.1. The setting for $\tilde Q$ is the same as that in the simulations. From these results, we can see that the results of $\tilde Q$ indicate that no serial correlation exists in the residuals. It is not surprise by noting that $\tilde Q$ suffers from the undersized issue. On the other hand, both EL and WeL suggest rejecting some of the null hypotheses when $m = 2$, and EL suggests rejecting most of them when $m = 6$. Considering the good finite performance of EL as indicated in simulations, we may conclude that the results  fitted by \emph{auto.arima.R} sound good. Note that based on the testing results of $\tilde Q$, it seems difficult to obtain such a conclusion.


\subsection{The PM2.5 in different cities}

In our second application, we consider testing the possibility of existing serial correlation in residuals when using the ARMA model to fit the daily PM2.5 data. The PM2.5 data are taken from \url{http://www.weather.com.cn/}. Many researchers considered fitting these data by using the ARMA model; see, e.g., \cite{cheng2019hybrid,wang2017novel,zhang2018trend}. Some of them found that there may exist ARCH effect in the PM2.5 data \citep{yao2022role}. Motivated by this, we also fit these datasets by using the ARMA-GARCH models based on \emph{auto.arima.R} and then test the possibility of existing serial correlation in the estimated residuals.

Since they are daily data, most of the related QQ-plots deviate from the diagonal line $y = x$, implying that their variances may possibly be quite large. Here, we do not present the QQ-plots for all these datasets in order to save space; see Figure 2 for details. The values of $p$, $q$ are selected automatically by \emph{auto.arima.R}. We then test $\mathcal{H}_0$ with three methods mentioned above. Their results are summarized in Table 4.2 for $m = 2$. From these results, it is easy to check that the $\tilde Q$ statistic rarely rejects the null hypothesis, while the EL rejects the null hypothesis for almost all datasets. Compared to $\tilde Q$ and EL, WeL appears to have a relatively reasonable rejection, considering that $\tilde Q$ and EL suffer from a significant size distortion problem as indicated in the simulations.

Note that the true conditional variances of the daily datasets may possibly tend to infinite, whereas when the true variance tends to infinite, the method in \emph{auto.arima.R} performs poorly in selecting the order of $p$, $q$ owning to its lack of consideration of the effect of infinite variance \citep{hyndman2008automatic}. In this sense, it is reasonable to consider that some of the residuals fitted by \emph{auto.arima.R} may show serial correlation because \emph{auto.arima.R} may select wrong $p$ or/and $p$ in some situations. It seems that this can not be reflected by the $\tilde Q$ and EL tests. 


\begin{figure}[h]
\flushleft
\includegraphics[scale = 0.30]{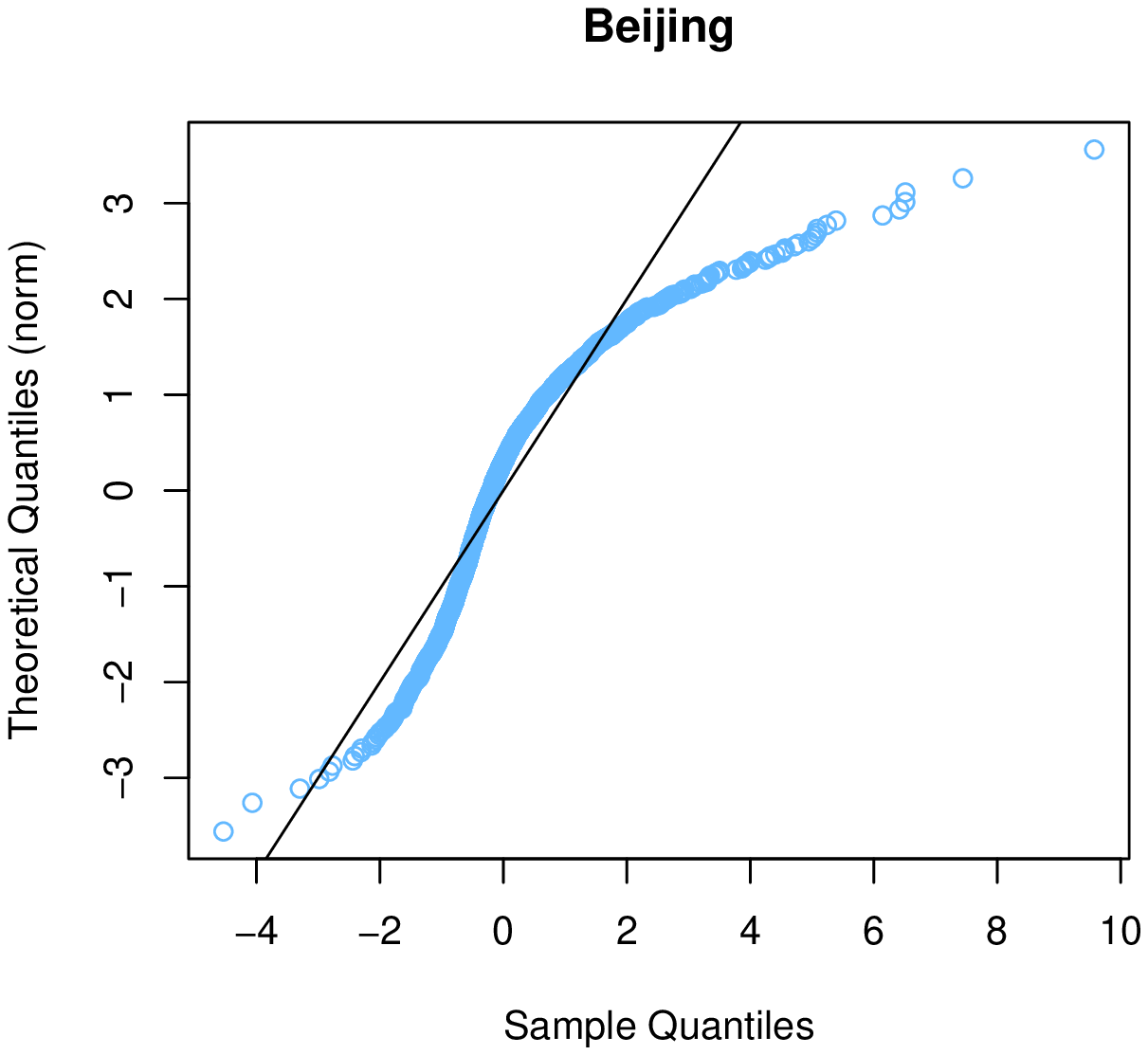}
\includegraphics[scale = 0.30]{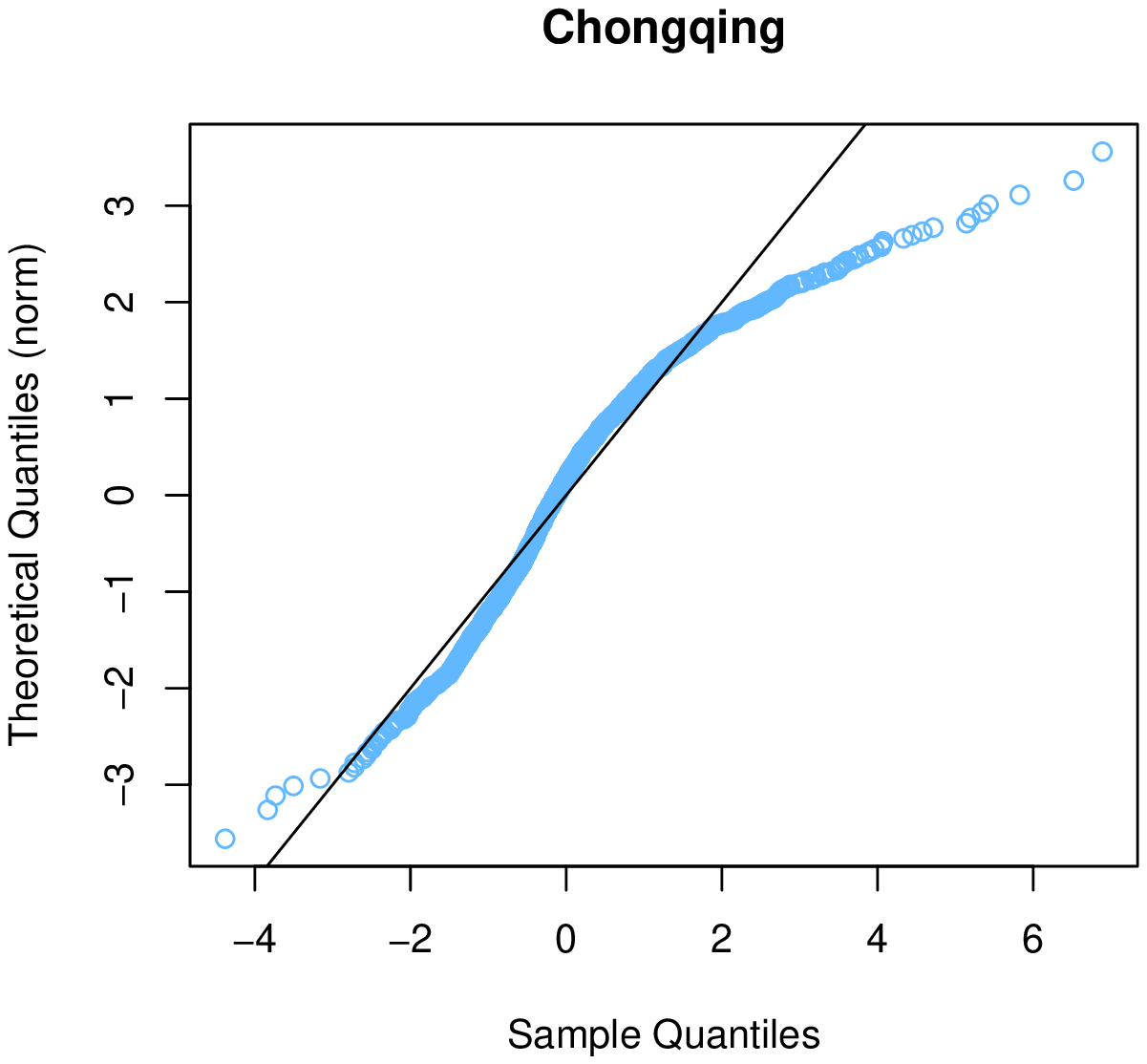}
\includegraphics[scale = 0.30]{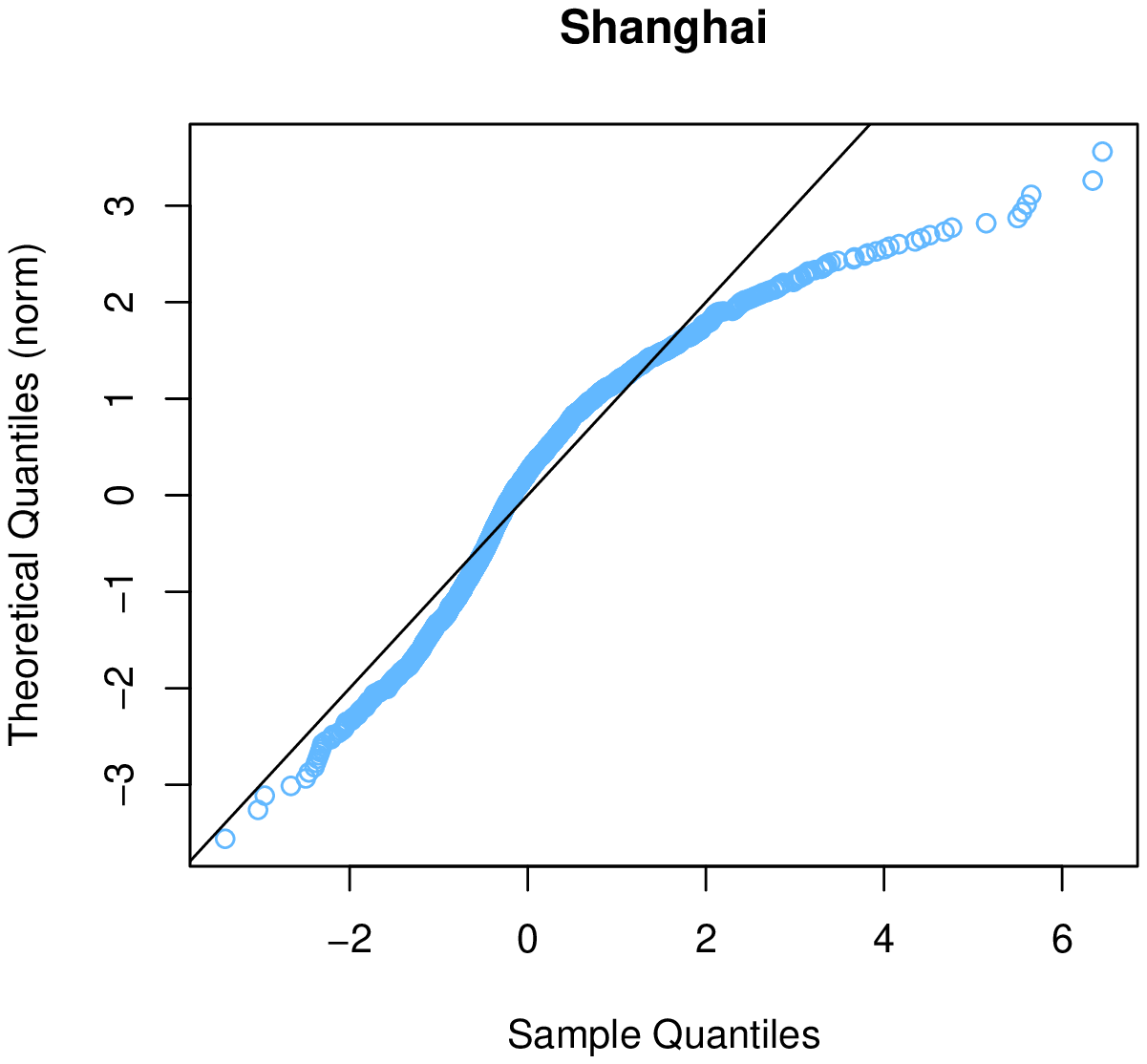}
\includegraphics[scale = 0.30]{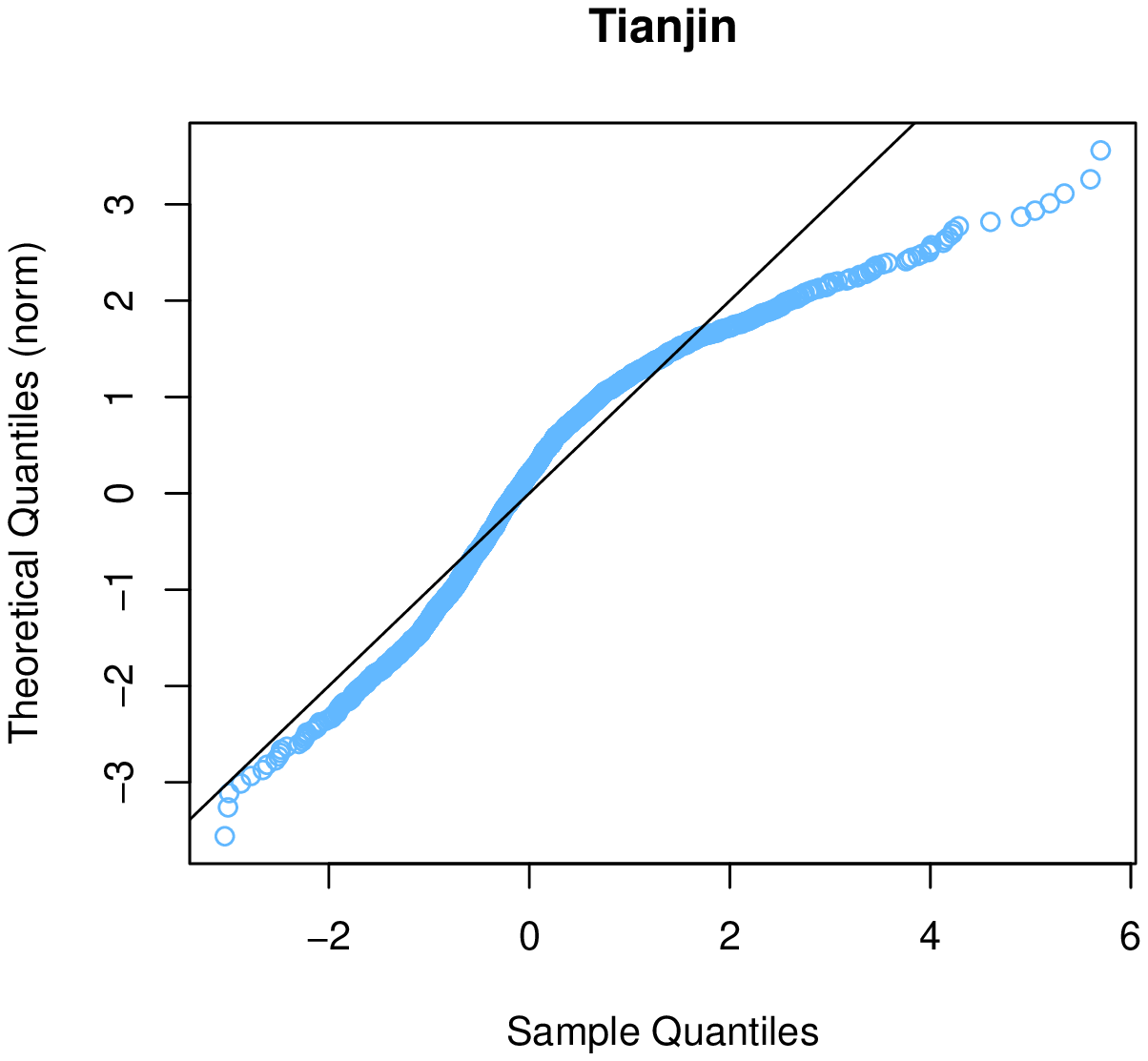}
\includegraphics[scale = 0.30]{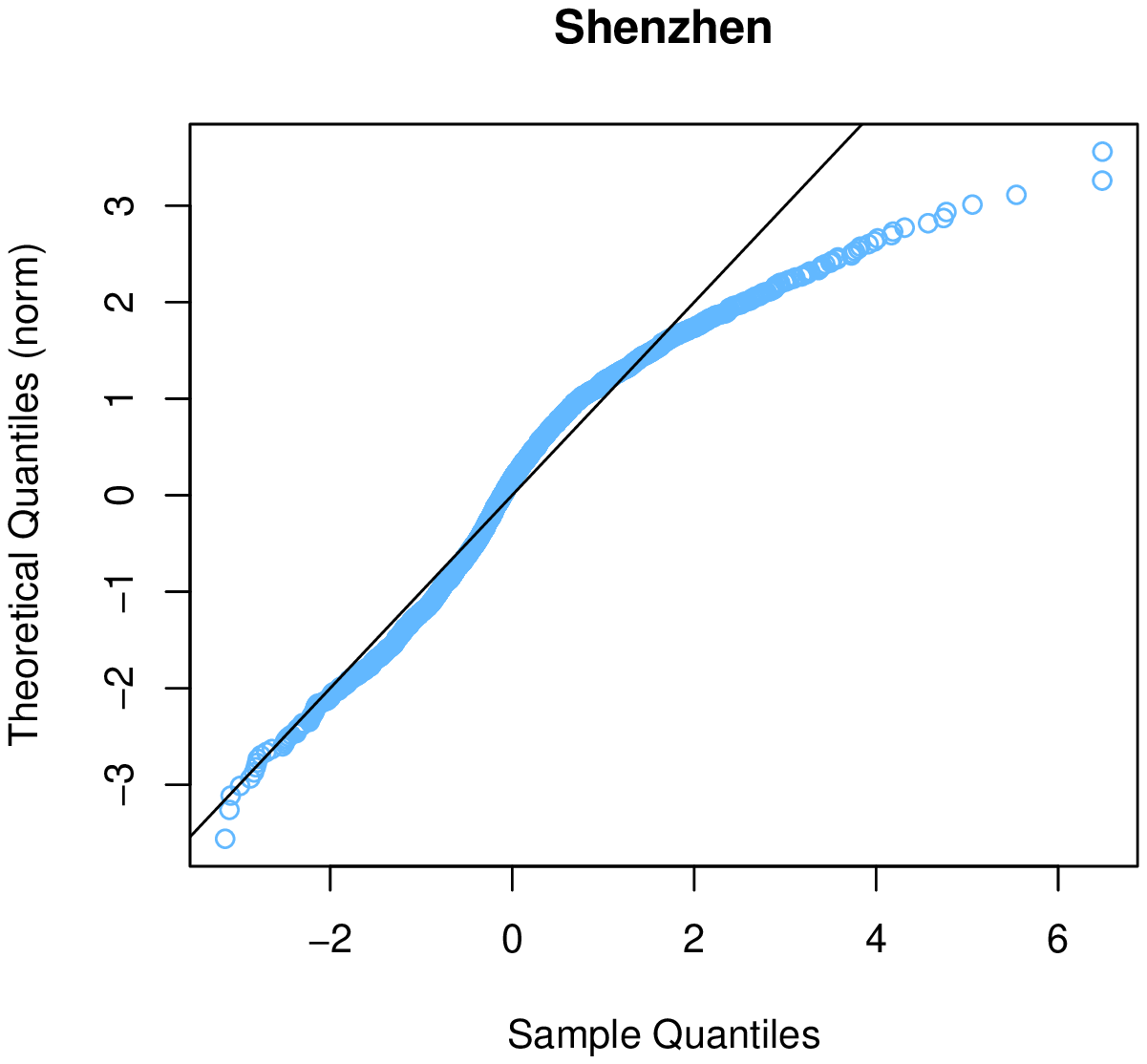}
\includegraphics[scale = 0.30]{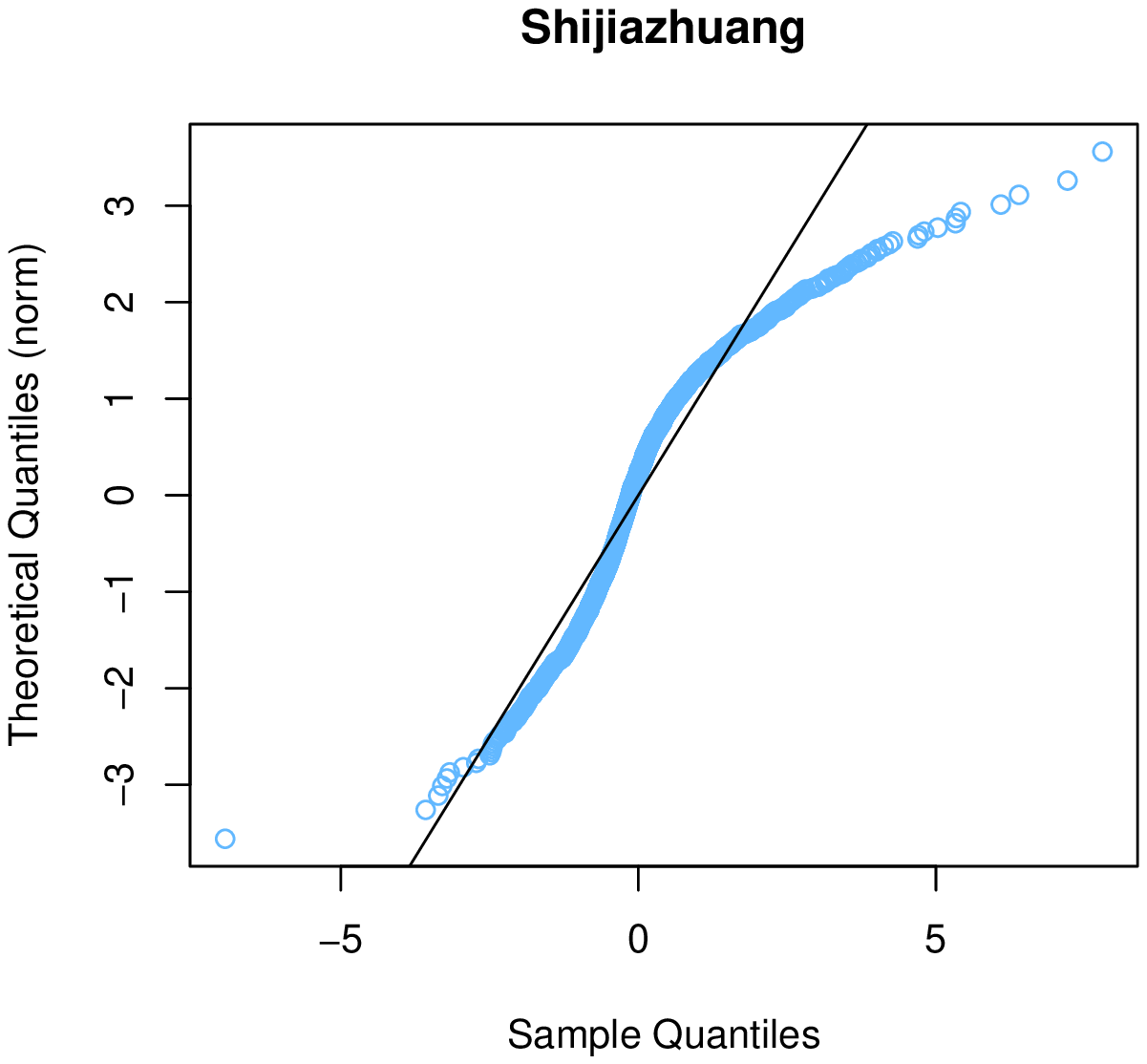}
\includegraphics[scale = 0.30]{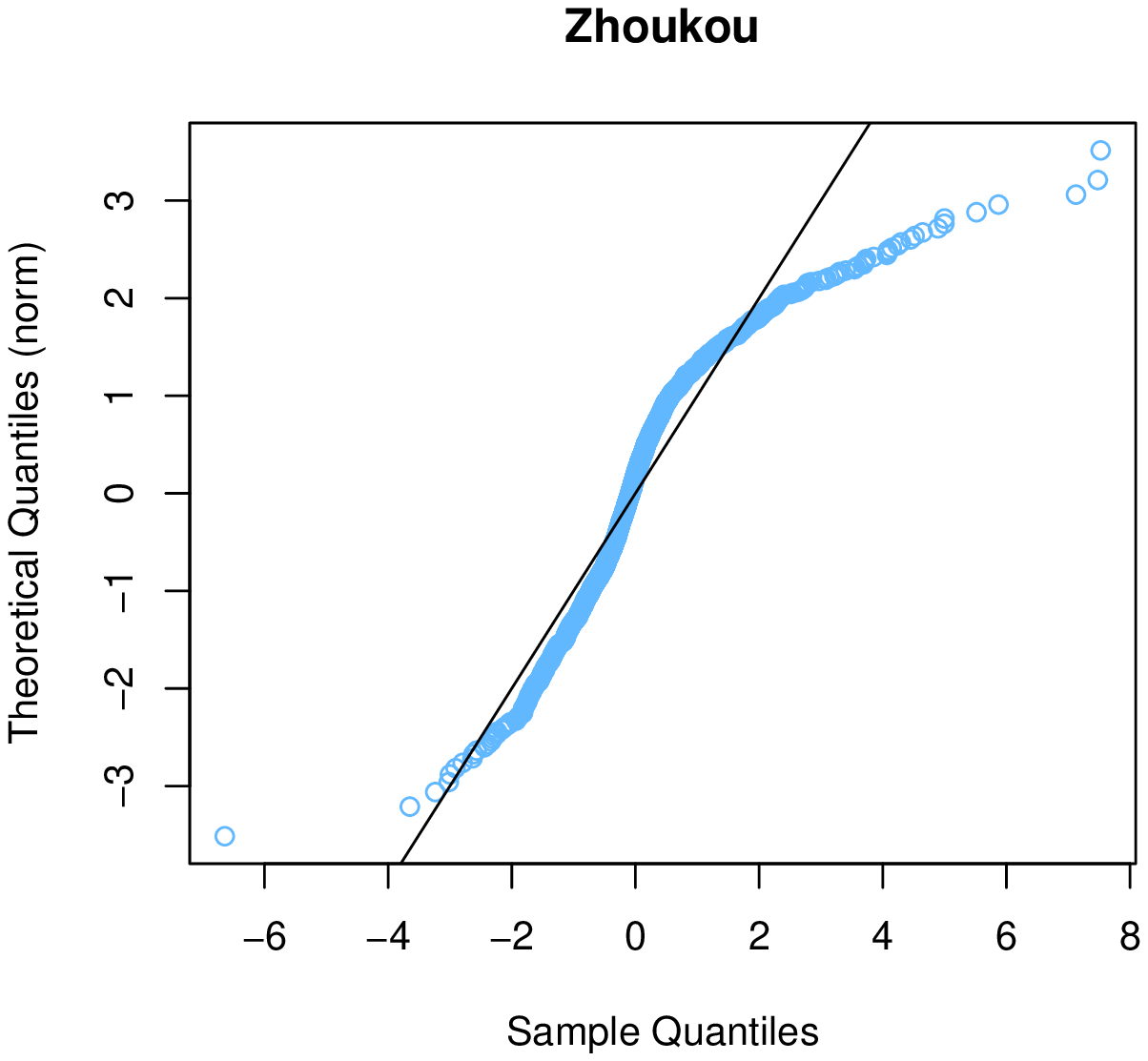}
\includegraphics[scale = 0.30]{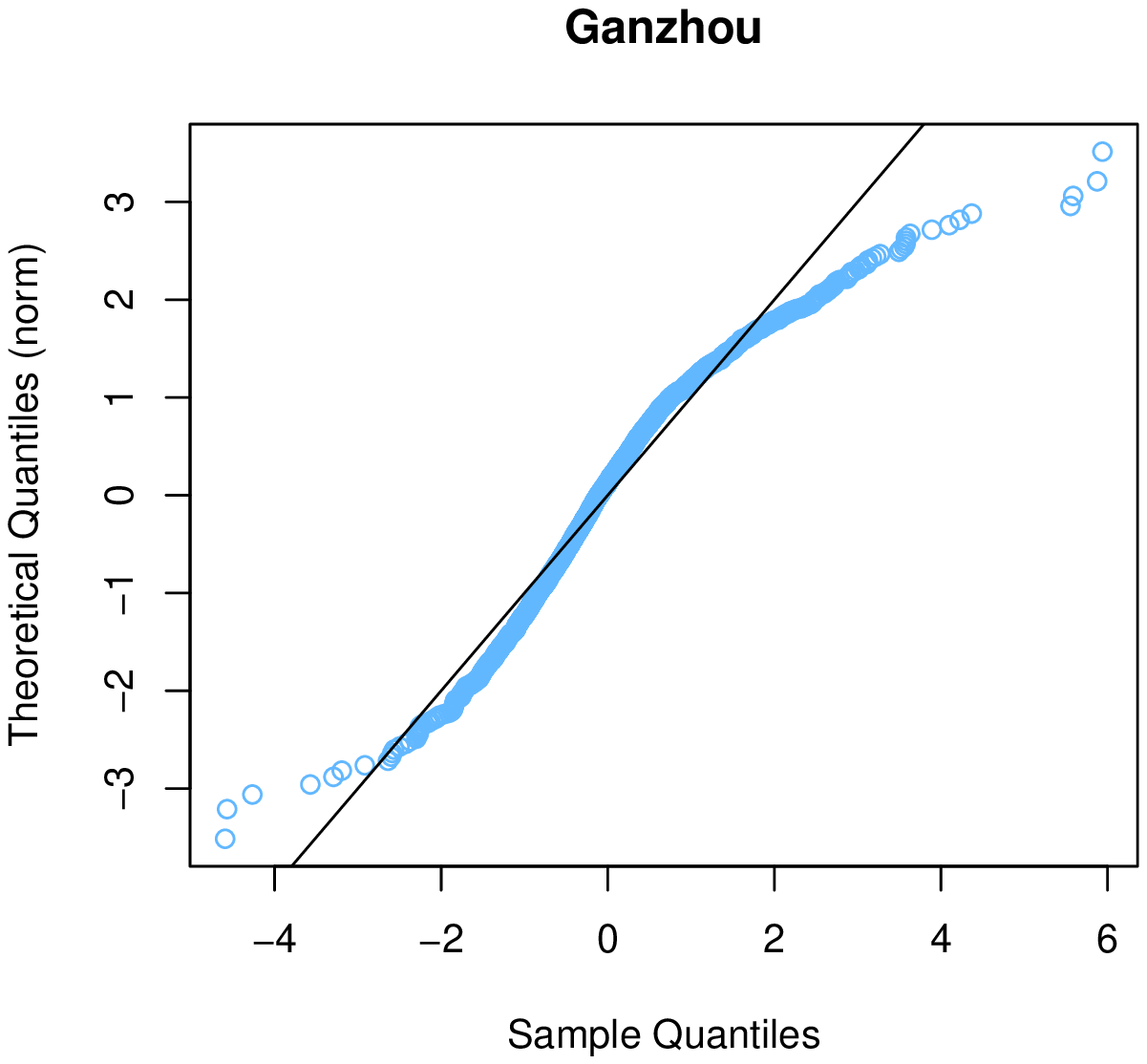}
\caption{QQ-plots for residuals of the daily PM2.5 data of eight cities.}
\label{Fig:Auto2}
\end{figure}

\begin{table*}
\setlength{\belowcaptionskip}{.2cm}  \textbf{TABLE 4.2}\\
 \centering The $p$-values of different tests with the daily PM2.5 data with $m = 2$.
\renewcommand{\arraystretch}{0.8} 
\begin{threeparttable}
 \setlength{\tabcolsep}{4mm}{
 \begin{tabular}{rrclllllll}
  \toprule
   \quad          & Cities &Time                 & ~~$\tilde Q$   &~~EL     &~WeL   \\
    \hline
    \quad & Chongqing & $2013.10-2021.04$ & 0.1474 & ${0.0016}^{\ast \ast \ast}$ & 0.8223 \\
    \quad & Xiamen    & $2013.10-2021.04$ & 0.4860 & ${0.0000}^{\ast \ast \ast}$ & ${0.0492}^{\ast \ast}$ \\
    \quad & Suzhou    & $2015.01-2021.04$ & 0.8091 & ${0.0000}^{\ast \ast \ast}$ & 0.9975 \\
    \quad & Liuan     & $2015.01-2021.04$ & 0.3916 & ${0.0000}^{\ast \ast \ast}$ & 0.7435 \\
    \quad & Maanshan  & $2014.01-2021.04$ & 0.4934 & ${0.0000}^{\ast \ast \ast}$ & 0.8317 \\
    \quad & Tongling  & $2015.01-2021.04$ & 0.1402 & ${0.0000}^{\ast \ast \ast}$ & 0.9623 \\
    \quad & Hangzhou  & $2013.10-2021.04$ & ${0.0490}^{\ast}$ & ${0.0005}^{\ast \ast \ast}$ &${0.0000}^{\ast \ast \ast}$  \\
    \quad & Anyang    & $2014.01-2021.04$ & 0.3370 & ${0.0000}^{\ast \ast \ast}$ & 0.9008 \\
    \quad & Hebi      & $2015.01-2021.04$ & 0.3070 & ${0.0000}^{\ast \ast \ast}$ & 0.8794 \\
    \quad & Jiaozuo   & $2014.01-2021.04$ & 0.5782 & ${0.0000}^{\ast \ast \ast}$ & 0.9810 \\
    \quad & Baoshan   & $2015.01-2021.04$ & 0.8771 & ${0.0000}^{\ast \ast \ast}$ & 0.4814 \\
    \quad & Ningbo    & $2013.10-2021.04$ & 0.2391 & ${0.0000}^{\ast \ast \ast}$ & ${0.0000}^{\ast \ast \ast}$ \\
    \quad & Shaoxing  & $2013.10-2021.04$ & 0.1159 & ${0.0584}^{\ast}$& 0.9701 \\
    \quad & Taizhou   & $2013.10-2021.04$ & 0.1032 & ${0.0001}^{\ast \ast \ast}$ & ${0.0000}^{\ast \ast \ast}$ \\
    \quad & Wenzhou   & $2013.10-2021.04$ & 0.1422 & ${0.0000}^{\ast \ast \ast}$ & ${0.0000}^{\ast \ast \ast}$ \\
    \quad & Yiwu      & $2014.01-2021.04$ & 0.2324 & ${0.0000}^{\ast \ast \ast}$ & ${0.0000}^{\ast \ast \ast}$ \\
    \quad & Zhoushan  & $2013.10-2021.04$ & 0.5299 & ${0.0000}^{\ast \ast \ast}$ & ${0.0118}^{\ast \ast}$\\
    \quad & Fuyang    & $2014.01-2021.04$ & ${0.0825}^{\ast}$ & ${0.0000}^{\ast \ast \ast}$ & 0.9168 \\
    \quad & Aba       & $2015.01-2021.04$ & 0.3172 & 0.8954 & ${0.0000}^{\ast \ast \ast}$ \\
    \quad & Chengdu   & $2013.10-2021.04$ & 0.9655 & ${0.0000}^{\ast \ast \ast}$ & ${0.0000}^{\ast \ast \ast}$ \\
    \bottomrule
    \end{tabular}}
     \begin{tablenotes}
        \footnotesize
        \item Significance levels: $ ^{\ast} p < 0.1, ^{\ast \ast} p < 0.05, ^{\ast \ast \ast} p< 0.01.$
      \end{tablenotes}
  \end{threeparttable}
\end{table*}

\section{Concluding discussions}
\label{Concluding}

In this paper, we considered the issue of diagnostic checking of AMAR models with a GARCH error by using the empirical likelihood. It turns out that the proposed log-empirical likelihood functions converge to a standard chi-squared distribution asymptotically. Since the empirical likelihood function does not involve the estimation of unknown variance, the new statistics do not need to estimate the GARCH parameters. We also compare the new method with the $\tilde{Q}$ statistic discussed in \cite{zhu2016bootstrapping}. It turns out the empirical likelihood-based methods perform better than $\tilde{Q}$ especially when the model has low persistence, and are both computationally easy. Note that since a weighted technique is employed to reduce the moment effect of $\sigma_t$, the weighted empirical likelihood statistic suffers a little power loss when the underlying model variance is finite.

\vskip 0.1 in
\section*{Appendix: Proofs of the main results}
\label{applications}

In this appendix, we provide the detailed proofs for the main results. Since the proof of Theorem 1 is like that of Theorem 2. We only prove Theorem 2. Without confusion, denote $\bmtht_0$ as the true value of $\bmtht$, and $\mathcal{F}_{t}$ as the sigma field generated by $\{\eta_s: s \le t\}$, and let
\begin{eqnarray*}
  \tilde{\bm Z}_{t}(\bmtht, 0) := \begin{pmatrix}
    \tilde{\bm Z}_{t,1}(\bmtht, 0)\\
    \tilde{\bm Z}_{t,2}(\bmtht, 0)
  \end{pmatrix},
\end{eqnarray*}
where $\tilde{\bm Z}_{t,1}(\bmtht, 0) = w_{t-1}^{-2} \varepsilon_t(\bm\theta) \partial \varepsilon_t(\bm\theta) / {\partial \bm\theta}$, and $\tilde{\bm Z}_{t,2}(\bmtht, 0) = (w_{t-1}^{-1}w_{t-2}^{-1}\varepsilon_t(\bm\theta) \varepsilon_{t-1}(\bm\theta), \cdots,w_{t-1}^{-1}$\\$ w_{t-m-1}^{-1}\varepsilon_t(\bm\theta) \varepsilon_{t-m}(\bm\theta))^\top$, for $t = m+1, 2, \cdots, n$.

The following lemmas are useful in proving Theorem 2.

\begin{lemma}\label{lem:01}
  Suppose the same conditions of Theorem \ref{th:002} holds. Then, there exist a constant $\rho \in (0, 1)$, a constant $C > 0$, and a neighborhood $\Theta_0$ such that
  \begin{eqnarray*}
    &\sup_{\bmtht \in\Theta_0} |\varepsilon_t(\bmtht)| \le C \xi_{\rho, t - 1}, \quad \sup_{\bmtht \in\Theta_0} \left\|\frac{\partial\varepsilon_t(\bmtht)}{\partial \bmtht}\right\| \le C \xi_{\rho, t - 1},
  \end{eqnarray*}
  and
  \begin{eqnarray*}
    &\sup_{\bmtht \in\Theta_0} \left\|\frac{\partial^2\varepsilon_t(\bmtht)}{\partial \bmtht \partial \bmtht^\top}\right\| \le C \xi_{\rho, t - 1},
  \end{eqnarray*}
  where $\xi_{\rho, t - 1}$ is defined in Condition \textbf{(C3)}, and $\|A\|^2 = \text{trace}(A^\top A)$ for a given matrix $A$.
\end{lemma}

\begin{proof}
  This lemma is adopted from \cite{ling2007self}. We omit the details.
\end{proof}

\begin{lemma}
  \label{lem:02}
  Let $\mathcal{B}_0 = \{\bmtht: \|\bmtht - \bmtht_0\| \le \frac{C}{\sqrt{n}}\}$ for some positive $C$. Then, under the same conditions of Theorem \ref{th:002}, as $n \to \infty$, we have uniformly for $\bmtht \in\mathcal{B}_0$ that:
  \begin{itemize}
    \item[(i).] $\max\limits_{m+1\le t\le n} \sup_{\bmtht \in \mathcal{B}_0} \|\tilde{\bm Z}_{t}(\bmtht, 0)\| = o_p(\sqrt{n})$;
    \item[(ii).] $\frac{1}{N} \sum_{t=m+1}^n \tilde{\bm Z}_{t}(\bmtht, 0) = \frac{1}{n} \sum_{t=m+1}^n \tilde{\bm Z}_{t}(\bmtht_0, 0) + O_p(\frac{1}{\sqrt n})$;
    \item[(iii).] $\frac{1}{N} \sum_{t=m+1}^n \tilde{\bm Z}_{t}(\bmtht, 0) \tilde{\bm Z}_{t}(\bmtht, 0)^\top = \tilde\Sigma  + o_p(1)$, where $\tilde \Sigma = E(\tilde{\bm Z}_{1}(\bmtht_0, 0)\tilde{\bm Z}_{1}(\bmtht_0, 0)^\top)$.
  \end{itemize}
\end{lemma}

\begin{proof}
  We first prove Part (i). Note that
  \begin{eqnarray*}
    \|\tilde{\bm Z}_{t}(\bmtht, 0)\| \le \|\tilde{\bm Z}_{t,1}(\bmtht, 0)\| + \sum_{l = 1}^m |\tilde{Z}_{t,p+q+l}(\bmtht, 0)|.
  \end{eqnarray*}
  By the proof of (i) in Lemma 2 of \cite{ma2021test},
  \begin{eqnarray*}
    \max\limits_{m+1\le t\le n} \sup_{\bmtht \in \mathcal{B}_0} \|\tilde{\bm Z}_{t,1}(\bmtht, 0)\| = o_p(\sqrt n),
  \end{eqnarray*}
  For $\tilde{Z}_{t,p+q+l}(\bmtht, 0)$, $l \in \{ 1, 2, \cdots, m\}$, note that
  \begin{eqnarray*}
    \sup_{\bmtht \in \mathcal{B}_0} |\tilde{Z}_{t,p+q+l}(\bmtht, 0)| &=& \sup_{\bmtht \in \mathcal{B}_0} |w_{t-1}^{-1} \varepsilon_{t}(\bmtht)w_{t-1-l}^{-1} \varepsilon_{t-l}(\bmtht)|\\
    &\le& C^2\underbrace{w_{t-1}^{-1}  \xi_{\rho, t - 1}}_{U_{t-1}} \underbrace{w_{t-1-l}^{-1} \xi_{\rho, t - 1 - l}}_{U_{t-1-l}},
  \end{eqnarray*}
  by following Lemma \ref{lem:01}. For any $\epsilon > 0$, by the Markov inequality and Cauchy-Schwarz inequality, it follows
  \begin{eqnarray*}
    &&P\left(\max_{m+1\le t\le n} U_{t-1}U_{t-1-l} \ge \sqrt{n} \epsilon\right)\\
    &\le& \sum_{t=m+1}^n P\left(U_{t-1}U_{t-1-l} \ge \sqrt{n} \epsilon\right)\\
    &\le& \frac{1}{n\sqrt{n}^{\delta/2} \epsilon^{2+\delta/2}} \sum_{t=m+1}^n E\left(U_{t-1}^{2+\delta/2} U_{t-1-l}^{2+\delta/2}\right)\\
    &\le& \frac{1}{\sqrt{n}^{\delta/2} \epsilon^{2+\delta/2}} \left\{\frac{1}{n} \sum_{t=m+1}^n \sqrt{E\left(U_{t-1}^{4+\delta})\right) E\left(U_{t-1-l}^{4+\delta}\right)}\right\} \\
    &\to& 0,
  \end{eqnarray*}
  as $n \to \infty$, based on Condition \textbf{(C3)}. This implies (i).

  For (ii), since the proof of
  \begin{eqnarray*}
    \frac{1}{N} \sum_{t=m+1}^n \tilde{\bm Z}_{t, 1}(\bmtht, 0) = \frac{1}{N} \sum_{t=m+1}^n \tilde{\bm Z}_{t, 1}(\bmtht_0, 0) + O_p(\frac{1}{\sqrt n})
  \end{eqnarray*}
  can be found in Lemma 2 of \cite{ma2021test}, we only need to show
  \begin{eqnarray*}
    \frac{1}{N} \sum_{t=m+1}^n \tilde{\bm Z}_{t, 2}(\bmtht, 0) = \frac{1}{N} \sum_{t=m+1}^n \tilde{\bm Z}_{t, 2}(\bmtht_0, 0) + O_p(\frac{1}{\sqrt n}).
  \end{eqnarray*}
  Note that
  \begin{eqnarray*}
    &&|w_{t-1}^{-1} \varepsilon_{t}(\bmtht)w_{t-1-l}^{-1} \varepsilon_{t-l}(\bmtht) - w_{t-1}^{-1} \varepsilon_{t}(\bmtht_0)w_{t-1-l}^{-1} \varepsilon_{t-l}(\bmtht_0)|\\
    &\le& \underbrace{|w_{t-1}^{-1} (\varepsilon_{t}(\bmtht) - \varepsilon_{t}(\bmtht_0))w_{t-1-l}^{-1} \varepsilon_{t-l}(\bmtht)|}_{V_{t,1}} \\
    &&- \underbrace{|w_{t-1}^{-1} \varepsilon_{t}(\bmtht_0)w_{t-1-l}^{-1} (\varepsilon_{t-l}(\bmtht) - \varepsilon_{t-l}(\bmtht_0))|}_{V_{t,2}}.
  \end{eqnarray*}
  A simple derivation leads to that
  \begin{eqnarray*}
    &&\sup_{\bmtht \in \mathcal{B}_0} V_{t,1}\\
    &\le&  \sup_{\bmtht \in \mathcal{B}_0} \left\{|w_{t-1}^{-1} w_{t-1-l}^{-1} \varepsilon_{t-l}(\bmtht)| \left\|\frac{\partial \varepsilon_{t-l}(\bmtht^*)}{\partial \bmtht}\right\| \|\bmtht - \bmtht_0\|\right\}\\
    &\le& \frac{C^2C_0}{\sqrt{n}} w_{t-1}^{-1} w_{t-1-l}^{-1} \xi_{\rho, t - 1} \xi_{\rho, t - 1 - l},
  \end{eqnarray*}
  where $\bmtht^*$ lies between $\bmtht$ and $\bmtht_0$. This implies as $n\to \infty$ that
  \begin{eqnarray*}
    \frac{1}{N} \sum_{t=m+1}^n \sup_{\bmtht \in \mathcal{B}_0} V_{t,1} = O_p\left(\frac{1}{\sqrt n}\right),
  \end{eqnarray*}
  under Condition \textbf{(C3)}. Similarly, we can show
  \begin{eqnarray*}
    \frac{1}{N} \sum_{t=m+1}^n \sup_{\bmtht \in \mathcal{B}_0} V_{t,2} = O_p\left(\frac{1}{\sqrt n}\right),\quad \text{as } n\to \infty.
  \end{eqnarray*}
  Hence, (ii) follows.

  The proof of (iii) follows as similar fashion to that of (ii). We omit the details.
\end{proof}

\begin{lemma}\label{lem:04}
  Under the same conditions of Theorem 2, we have, as $n \to \infty$,
  \begin{eqnarray*}
    &&\frac{1}{\sqrt N} \sum_{t=m+1}^n \tilde{\bm Z}_{t}(\bmtht_0, 0) \overset{d}{\longrightarrow} N(0, \tilde\Sigma),~~~\text{and }\\
    && \frac{1}{ N} \sum_{t=m+1}^n \tilde{\bm Z}_{t}(\bmtht_0, 0) \tilde{\bm Z}_{t}^\top(\bmtht_0, 0) \overset{p}{\longrightarrow} \tilde\Sigma.
  \end{eqnarray*}
\end{lemma}

\begin{proof}
  It follows from the first part of Lemma 3 in \cite{ma2021test} that $\frac{1}{\sqrt N} \sum_{t=m+1}^n \tilde{\bm Z}_{t, 1}(\bmtht_0, 0)$ is asymptotically normally distributed. Then, it suffices to show that, as $n \to \infty$,
  \begin{eqnarray}\label{eqn:cross}
    &&\frac{1}{N} \sum_{t=m+1}^n E(\tilde{\bm Z}_{t, 1}(\bmtht_0, 0)w_{t-1}^{-1} w_{t-1-l}^{-1} \varepsilon_{t}(\bmtht_0) \varepsilon_{t-l}(\bmtht_0)|\mathcal{F}_{t-1})\\\nonumber &&\overset{p}{\longrightarrow} \lim_{t\to\infty} E\left(\sigma_t^2 \frac{\partial \varepsilon_{t}(\bmtht_0)}{\partial \bmtht}  w_{t-1}^{-3} w_{t-1-l}^{-1} \varepsilon_{t-l}(\bmtht_0)\right),
  \end{eqnarray}
  for $l = 1, 2, \cdots, m$, and
  \begin{eqnarray}\label{eqn:Zt2}
    &&\frac{1}{\sqrt N} \sum_{t=m+1}^n \tilde{\bm Z}_{t, 2}(\bmtht_0, 0) \\\nonumber &&\overset{d}{\longrightarrow} N(0, E(\tilde{\bm Z}_{1, 2}(\bmtht_0, 0)\tilde{\bm Z}_{1, 2}(\bmtht_0, 0)^\top)).
  \end{eqnarray}

  Note that
  \begin{eqnarray*}
    &&E(\tilde{\bm Z}_{t, 1}(\bmtht_0, 0)w_{t-1}^{-1} w_{t-1-l}^{-1} \varepsilon_{t}(\bmtht_0) \varepsilon_{t-l}(\bmtht_0)|\mathcal{F}_{t-1})\\
    &&= E(\varepsilon_{t}^2(\bmtht_0) \frac{\partial \varepsilon_{t}(\bmtht_0)}{\partial \bmtht}  w_{t-1}^{-3} w_{t-1-l}^{-1} \varepsilon_{t-l}(\bmtht_0)|\mathcal{F}_{t-1})\\
    && = \sigma_t^2 \frac{\partial \varepsilon_{t}(\bmtht_0)}{\partial \bmtht}  w_{t-1}^{-3} w_{t-1-l}^{-1} \varepsilon_{t-l}(\bmtht_0).
  \end{eqnarray*}
  We obtain \eqref{eqn:cross} under Conditions \textbf{(C1)} and \textbf{(C3)} based on the weak law of large numbers for a martingale difference series given in \cite{hall2014martingale} and the stationarity of $\{\sigma_t^2\}$, $\{X_t\}$, and $\{w_{t}\}$.

  For \eqref{eqn:Zt2}, let $W_t = \bm a^\top \tilde{\bm Z}_{t, 2}(\bmtht_0, 0)$ with $\bm a$ being an any given $m$-dimensional nonzero vector. Then, it is easy to check that $E(W_t|\mathcal{F}_{t-1}) = 0$, for any $t = 1, 2, \cdots, n$. That is, $\{W_t\}$ is a martingale difference sequence.

  Next, note that
  \begin{eqnarray}\label{eqn:varzt2}
    &&\frac{1}{N}\sum_{t=m+1}^n E(W_t^2|\mathcal{F}_{t-1})\\\nonumber
     &=& \bm a^\top \frac{1}{N}\sum_{t=m+1}^n E\left(\tilde{\bm Z}_{t, 2}(\bmtht_0, 0)\tilde{\bm Z}_{t, 2}(\bmtht_0, 0)^\top|\mathcal{F}_{t-1}\right) \bm a.
  \end{eqnarray}
  For any $1 \le i, j \le m$, since by Condition \textbf{(C3)} and the Cauchy-Schwarz inequality,
  \begin{eqnarray*}
    &&\left|\frac{1}{N}\sum_{t=m+1}^n w_{t-1}^{-2}\varepsilon_t^2(\bmtht_0)w_{t-1-i}^{-1}\varepsilon_{t-i}(\bmtht_0) w_{t-1-j}^{-1}\varepsilon_{t-j}(\bmtht_0)\right|\\
    &\le& \frac{1}{N}\sum_{t=m+1}^n \Big(\frac{1}{2}w_{t-1}^{-4}\varepsilon_t^4(\bmtht_0) + \frac{1}{4}w_{t-1-i}^{-4}\varepsilon_{t-i}^4(\bmtht_0) + \\ &&~~~~~~~~~~~~~~
    \frac{1}{4}w_{t-1-j}^{-4}\varepsilon_{t-j}^4(\bmtht_0)\Big)\\
    &\le& \frac{1}{N}\sum_{t=m+1}^n \Big(\frac{1}{2}w_{t-1}^{-4} \xi_{\rho, t-1}^4 + \frac{1}{4}w_{t-1-i}^{-4} \xi_{\rho, t-1-i}^4 \\ &&~~~~~~~~~~~~~~ \frac{1}{4}w_{t-1-j}^{-4} \xi_{\rho, t-1-j}^4\Big)\\
    &\overset{p}{\longrightarrow}& \lim_{t\to\infty} E(w_{t-1}^{-4} \xi_{\rho, t-1}^4),
  \end{eqnarray*}
  as $n \to \infty$, where `$\overset{p}{\longrightarrow}$' denotes the convergence in probability. Then, we may conclude that \eqref{eqn:varzt2} converges by the dominated convergence theorem and the weak law of large numbers for a martingale difference series given in \cite{hall2014martingale}.

  Furthermore, similar to the proof of \eqref{eqn:varzt2}, we can show that
  \begin{eqnarray*}
    \frac{1}{N}\sum_{t=m+1}^n E(W_t^2I(|W_t|\ge \sqrt n \epsilon)|\mathcal{F}_{t-1}) \overset{p}{\longrightarrow} 0, \text{ as } n \to \infty,
  \end{eqnarray*}
  for any positive $\epsilon > 0$. Finally, we complete the proof of this lemma by using the central limit theorem of martingale differences \citep{hall2014martingale}. This proves the first part.

  The second part follows a similar fashion. We omit the details.
\end{proof}


\begin{proof}[Proof of Theorem 2.]
  Based on Lemmas 2-3, the following proof is similar to that of Theorem 1 in \cite{ma2021test}.

  Put $\bm\theta = \bm\theta_0 + \frac{\bm u}{\sqrt n}$ for some $(p + q + 1)$-dimensional vector $\bm u$. Define
  \begin{eqnarray*}
    h(\bm\theta, \bm\lambda) = \frac{1}{N} \sum_{t=m+1}^N \frac{\tilde{\bm Z}_{t}(\bmtht, 0)}{1 + \bm\lambda^\top \tilde{\bm Z}_{t}(\bmtht, 0)},
  \end{eqnarray*}
  where $\bm\lambda$ is the solution to $h(\bm\theta, \bm\lambda) = 0$ for given $\bm\lambda$.

  Write $\bm\theta = \rho \bm v$ with $\|\bm v\| = 1$. Note that
  \begin{eqnarray*}
    0 = \|h(\bm\theta, \bm\lambda)\| \ge |\bm v^\top h(\bm\theta, \bm\lambda)| = \left|\frac{1}{N} \sum_{t=m+1}^N \frac{\bm v^\top\tilde{\bm Z}_{t}(\bmtht, 0)}{1 + \rho \bm v^\top \tilde{\bm Z}_{t}(\bmtht, 0)}\right|.
  \end{eqnarray*}
  Then, by a standard proof as that in \cite{owen2001empirical} we can show that $\bm\lambda = O_p(\frac{1}{\sqrt N})$, and
  \begin{eqnarray*}
    \bm\lambda = T_n^{-1}(\bm\theta,0) \left(\frac{1}{N}\sum_{t=m+1}^N \tilde{\bm Z}_{t}(\bmtht, 0)\right) + o_p\left(\frac{1}{\sqrt N}\right),
  \end{eqnarray*}
  uniformly for $\bm\theta \in \mathcal{B}_0$ based on Lemma \ref{lem:02}, where $T_n(\bm\theta,0) = \frac{1}{N}\sum_{t=m+1}^N \tilde{\bm Z}_{t}(\bmtht, 0)\tilde{\bm Z}_{t}^\top(\bmtht, 0)$. Using this, we can further derive by the Taylor expansion and Lemma \ref{lem:02} that
  \begin{eqnarray*}
    &&-2\log(\tilde L(\bmtht, 0))\\
    &=& 2\log(1 + \bm\lambda^\top \tilde{\bm Z}_{t}(\bmtht, 0))\\
    &=& 2\bm\lambda^\top (\sum_{t=m+1}^N \tilde{\bm Z}_{t}(\bmtht, 0)) - N\bm\lambda^\top T_n(\bm\theta, 0) \bm\lambda \\
    && + \frac{2}{3!} \sum_{t=m+1}^N \frac{1}{(1 + \xi_t^*)^2}(\bm\lambda^\top \tilde{\bm Z}_{t}(\bmtht, 0))^3\\
    &=& S_n(\bmtht, 0)^\top T_n^{-1}(\bm\theta, 0) S_n(\bmtht, 0) + o_p(1)\\
    &=& S_n(\bmtht, 0)^\top \tilde\Sigma^{-1} S_n(\bmtht, 0) + o_p(1),
  \end{eqnarray*}
  uniformly for $\bm\theta \in \mathcal{B}_0$, where $|\xi_t^*| < |\bm\lambda^\top \tilde{\bm Z}_{t}(\bmtht, 0)|$, $S_n(\bmtht, 0) = \frac{1}{\sqrt N} \sum_{t=m+1}^N \tilde{\bm Z}_{t}(\bmtht, 0)$. Note that
  \begin{eqnarray*}
    &&\left|\sum_{t=m+1}^N \frac{1}{(1 + \xi_t^*)^2}(\bm\lambda^\top \tilde{\bm Z}_{t}(\bmtht, 0))^3\right|\\
    &\le& C \sum_{t=m+1}^N \|\bm\lambda\|^3 \|\tilde{\bm Z}_{t}(\bmtht, 0)\|^3 = o_p(1),
  \end{eqnarray*}
  uniformly for $\bm\theta \in \mathcal{B}_0$ based on Lemma \ref{lem:02} as $n \to \infty$.

  Furthermore, since $\bm\theta_0 \in \mathcal{B}_0$, we have as $n \to \infty$
  \begin{eqnarray*}
    -2\log(\tilde L(\bmtht_0, 0)) = S_n(\bmtht_0, 0)^\top \tilde\Sigma^{-1} S_n(\bmtht_0, 0) + o_p(1).
  \end{eqnarray*}
  That is,
  \begin{eqnarray}\label{eqn:diffL}
    &&-2\log(\tilde L(\bmtht, 0)) + 2\log(\tilde L(\bmtht_0, 0))\\
    &=& S_n(\bmtht, 0)^\top \tilde\Sigma^{-1} S_n(\bmtht, 0) - \nonumber\\
    &&~~~S_n(\bmtht_0, 0)^\top \tilde\Sigma^{-1} S_n(\bmtht_0, 0) + o_p(1).\nonumber
  \end{eqnarray}

  Note that for given $\bm\theta$, by the Taylor expansion and Lemmas \ref{lem:01}-\ref{lem:02}, we have
  \begin{eqnarray*}
    &&S_n(\bm\theta) - S_n(\bm\theta_0) \\
    &=& \frac{1}{\sqrt N} \sum_{t=m+1}^N \begin{pmatrix}
    \tilde{\bm Z}_{t,1}(\bmtht, 0) - \tilde{\bm Z}_{t,1}(\bmtht_0, 0)\\
    \tilde{\bm Z}_{t,2}(\bmtht, 0) - \tilde{\bm Z}_{t,2}(\bmtht_0, 0)
  \end{pmatrix}\\
  &=& \left(\frac{1}{N} \sum_{t=m+1}^N \begin{pmatrix}
    \frac{\partial(\tilde{\bm Z}_{t,1}(\bmtht_0, 0))}{\partial \bm\theta^\top} \\
    \frac{\partial(\tilde{\bm Z}_{t,2}(\bmtht_0, 0))}{\partial \bm\theta^\top}
  \end{pmatrix}\right) \sqrt N (\bm\theta - \bm\theta_0) + o_p(1)\\
  &=& E\begin{pmatrix}
    \frac{\partial(\tilde{\bm Z}_{t,1}(\bmtht_0, 0))}{\partial \bm\theta^\top} \\
    \frac{\partial(\tilde{\bm Z}_{t,2}(\bmtht_0, 0))}{\partial \bm\theta^\top}
  \end{pmatrix} \sqrt N (\bm\theta - \bm\theta_0) + o_p(1)\\
  &:=& \tilde{\Gamma}\sqrt N (\bm\theta - \bm\theta_0) + o_p(1),
  \end{eqnarray*}
  as $n \to \infty$. Hence, the minimizer, say $\hat{\bm\theta}$, of $-2\log(\tilde L(\bmtht, 0))$ with respect to $\bm\theta$ satisfies that
  \begin{eqnarray*}
    0 &=& \frac{-2\partial \log(\tilde L(\hat{\bm\theta}, 0))}{\partial \bm\theta} \\
    &=& 2\tilde{\Gamma} \tilde{\Sigma}^{-1} \tilde{\Gamma}^\top \sqrt N(\hat{\bm\theta} - \bm\theta_0) + 2\tilde{\Gamma} \tilde{\Sigma}^{-1} S_n(\bm\theta_0) + o_p(1).
  \end{eqnarray*}
  For given $\bm\theta$, let $\varrho = \sqrt N \|{\bm\theta} - \bm\theta_0\|$, and $\bm v =  \frac{{\bm\theta} - \bm\theta_0}{\|{\bm\theta} - \bm\theta_0\|}$. Then, it is easy to check that
  \begin{eqnarray*}
    &&\left\|\frac{-2\partial \log(\tilde L({\bm\theta}, 0))}{\partial \bm\theta}\right\|\\
    &\ge& \left|\bm v^\top \frac{-2\partial \log(\tilde L({\bm\theta}, 0))}{\partial \bm\theta}\right|\\
    &\ge& 2\varrho \bm v^\top\tilde{\Gamma} \tilde{\Sigma}^{-1} \tilde{\Gamma}^\top \bm v - 2|\bm v^\top\tilde{\Gamma} \tilde{\Sigma}^{-1} S_n(\bm\theta_0)| + o_p(1)\\
    &\overset{p}{\longrightarrow}& \infty, ~~~ \text{as } \varrho \to \infty,
  \end{eqnarray*}
  by noting that $\bm v^\top\tilde{\Gamma} \tilde{\Sigma}^{-1} \tilde{\Gamma}^\top \bm v = O_p(1)$ and $|\bm v^\top\tilde{\Gamma} \tilde{\Sigma}^{-1} S_n(\bm\theta_0)|$ $ = O_p(1)$ as $n \to \infty$. This shows $\hat{\bm\theta} \in \mathcal{B}_0$. Further combining with \eqref{eqn:diffL}, we obtain
  \begin{eqnarray*}
    \sqrt N(\hat{\bm\theta} - \bm\theta_0) = -(\tilde{\Gamma} \tilde \Sigma^{-1} \tilde{\Gamma}^\top)^{-1} (\tilde{\Gamma} \tilde \Sigma^{-1} S_n(\bm\theta_0)) + o_p(1),
  \end{eqnarray*}
  as $n \to \infty$.

  Finally, as in \cite{qin1994empirical}, we show that
  \begin{eqnarray*}
    &&\inf\left\{-2\log(\tilde L(\bm\theta, 0))\right\}\\
    &=&-2\log(\tilde L(\hat{\bm\theta}, 0))\\
    &=& S_n^\top(\bm\theta_0) (\tilde{\Sigma} - \tilde{\Gamma}^\top (\tilde{\Gamma} \tilde{\Sigma}^{-1} \tilde{\Gamma})^{-1}  \tilde{\Gamma}) S_n(\bm\theta_0) + o_p(1)\\
    &\overset{d}{\longrightarrow}& \chi_m^2,
  \end{eqnarray*}
  as $n \to \infty$. This completes the proof of this theorem.
\end{proof}

\bigskip

\bibliographystyle{agsm}
\bibliography{ARMAGARCH}

\end{document}